\documentclass[prl,preprint,showpacs,preprintnumbers,amsmath,amssymb]{revtex4}

\usepackage{graphicx}
\usepackage{dcolumn}
\usepackage{bm}
\usepackage{eurosym}
\usepackage{times}

\newcommand{\mltextt}{}
\newcommand{\nntext }{}
\newcommand{\ntekst}{}
\newcommand {\mmmmtext }{}
\newcommand{\mmmtext }{}
\newcommand{\ttext}{}

\newcommand{\mmtext}{  }
\newcommand{\mtext}{  }
\newcommand{\mltext}{}
\newcommand{\atxt}{ }

\newcommand{\prob}{{\mathbb P}}

\newcommand{\R}{{\mathbb R}}

\newcommand{\C}{{\mathbb C}}

\def\tilde{\widetilde}
\def\bfo{\begin{eqnarray*} }
\def\efo{\end{eqnarray*} }
\def\ba{\begin{eqnarray*} }
\def\ea{\end{eqnarray*} }
\def\beq{\begin{eqnarray}}
\def\eeq{\end{eqnarray}}

\def\det {\hbox{det}}

\def\e{\varepsilon}
\def\p{\partial}
\def\a{\alpha}
\def\k{{\bf k}}

\def\r{\rho}
\def\E{{\mathcal E}}

\def\p{\partial}

\def\R{\mathbb R}

\def\V{\mathcal V}

\def\x{{\bf x}}
\def\y{{\bf y}}

\newcommand{\radius}{R}

\def\etal{\emph{et al.}\,}

\def\sci{\emph{Science}\,}

\def\prln{\emph{Phys. Rev. Lett.}\,}

\begin{document}

\centerline{\large \bf Schr\"odinger's Hat: Electromagnetic, acoustic and}

\centerline{\large \bf  quantum  amplifiers via transformation optics}

\bigskip

\centerline{ Allan Greenleaf\,${}^{1,*}$, Yaroslav Kurylev${}^{2}$, Matti Lassas${}^{3}$ 
and Gunther Uhlmann${}^{4}$}

\centerline{\it  ${}^{1}$Dept.\ of Mathematics,
 University of Rochester, Rochester, NY 14627}

\centerline{\it   ${}^{2}$Dept.\ of Mathematical Sciences, University College London,  London,
WC1E 6BT, UK}

\centerline{\it  ${}^{3}$Dept.\ of Mathematics, University of  Helsinki, FIN-00014, Finland,}

\centerline{\it  ${}^{4}\hbox{Dept.\ of Mathematics, University of Washington, Seattle, WA 98195}$ and}

\centerline{\it Dept.\ of Mathematics, University of California, Irvine, CA 92697}

\centerline{\it ${}^*$Authors are listed in alphabetical order}

\centerline{(Version of  July 22, 2011)}


\begin{abstract} The advent of transformation optics  and metamaterials has 
made possible devices producing extreme effects on wave propagation.
Here we give   theoretical designs  for  devices,  \emph{Schr\"odinger hats}, acting as invisible concentrators of waves. These exist for any wave phenomenon modeled by either the Helmholtz or Schr\"odinger equations, e.g.,   polarized waves in EM, pressure waves in acoustics and matter waves in QM, and   occupy one part of a parameter space 
continuum of  wave-manipulating structures which also contains standard  transformation optics based cloaks, resonant cloaks and cloaked sensors. For EM and acoustic Schr\"odinger hats,  the resulting centralized wave is a localized excitation.
In QM, the result is a new charged quasiparticle,  a \emph{quasmon}, which causes conditional probabilistic illusions.
We discuss possible solid state implementations.
\end{abstract}

\vfill

\maketitle       

\newpage


Transformation optics  and metamaterials have 
made possible
devices producing  effects on wave propagation not seen in nature, including invisibility cloaks  for electrostatics \cite{GLU1,GLU2}, electromagnetism (EM) \cite{Le,PSS1,SchurigEtAl}, acoustics \cite{CummerSchurig,ChenChan,CummerEtAl} and quantum mechanics (QM) \cite{Zhang}; field rotators \cite{ChenRotate};  EM wormholes \cite{GKLUWorm}; and illusion optics \cite{ChanIllusion}, among many others.
The purpose of this paper is to give  theoretical designs  for  devices, which we  refer to as  \emph{Schr\"odinger hats}, acting as invisible concentrators, reservoirs  and amplifiers for waves. Schr\"odinger hats exist for any wave phenomenon modeled by either the Helmholtz or Schr\"odinger equation, whether in EM,  acoustics or QM.  Schr\"odinger hats (SH) occupy one part of a parameter space 
continuum of  wave-concentrating structures which also contains standard  transformation optics based cloaks and cloaked sensors \cite{GKLUSensor}. 
For EM and acoustic SH,  the resulting centralized wave is a localized excitation, which may be super-wavelength in scale; in QM, the SH produces a new quasiparticle, a \emph{quasmon}. Acoustic and EM   Schr\"odinger hats require negative index  materials, while highly oscillatory potentials are needed for QM hats. 
A SH   seizes a large fraction of  an incident wave, holding 
{\mtext and amplifying} it as a quasmon, 
while contributing only a negligible amount to scattering. 
 These devices
 are consistent with  the uncertainty principle, and we illustrate the concept by a QM version of  three card monte.
While a quantum Schr\"odinger hat is invisible to  one-particle scattering,  we show by effective potential theory \cite{Parr}  
that a SH acts as an  amplifier of  two-particle Coulomb interactions.
Such {\mtext  amplifiers} may be useful for   quantum  measurement and information processing.  The  similar  yet  less demanding acoustic and EM {\mtext  hats}  offer  equivalent effects, but existing metamaterials \cite{Shelby,LiChan,Lee,ChenChanSurvey}  make these designs more immediately realizable, 
allowing verification and further exploration of Schr\"odinger hats and quasmons.

 There are a number of  `paradoxes' in which the   laws of quantum mechanics  imply results that conflict with our intuition \cite{paradox}.
In this spirit, here we show that  the behavior of matter waves, as governed by Schr\"odinger's equation, combined with the virtual space/physical space paradigm of transformation optics, allows  one
to manipulate conditional probabilities in QM
and  create quantum illusions,  in which observed locations of particles
 differ from their actual values.

 \begin{figure}[htbp]
\begin{center}
\vspace{-8.5cm}
\hspace{-2cm}
\includegraphics[width=1.15\linewidth]{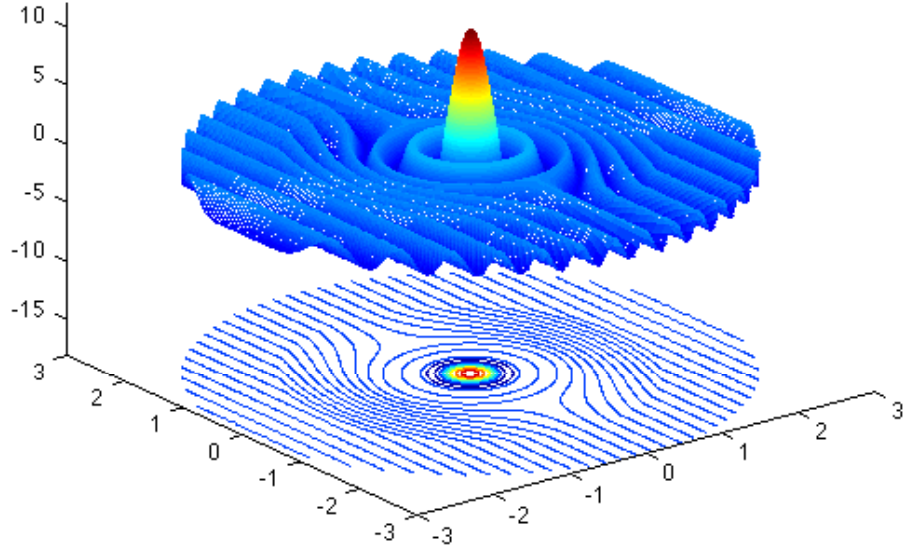}\hspace{-3.5cm}
\hspace{-8cm}
\end{center}
\vspace{-8cm}

\caption{{\bf A quasmon inside a Schr\"odinger hat.} 
The real part of the effective wave function $\psi^{sh}_{eff}(x,y,z)$  at the plane $z=0$ when a plane wave is incident  to a SH potential. 
By varying the design parameters, the concentration
of the wave inside the cloaked region can be made arbitrarily strong and the scattered field 
 arbitrarily small. The matter wave  is spatially localized, but conforms to the uncertainty principle, with the large gradient of $\psi$, visible as the steep slope  of the central peak,  concentrating the momentum  in a spherical shell in $p$-space. 
 }
\end{figure}

 \begin{figure}\label{fig three regimes}
 \begin{center}
 \vspace{-2cm}
 \includegraphics[width=.55\linewidth]{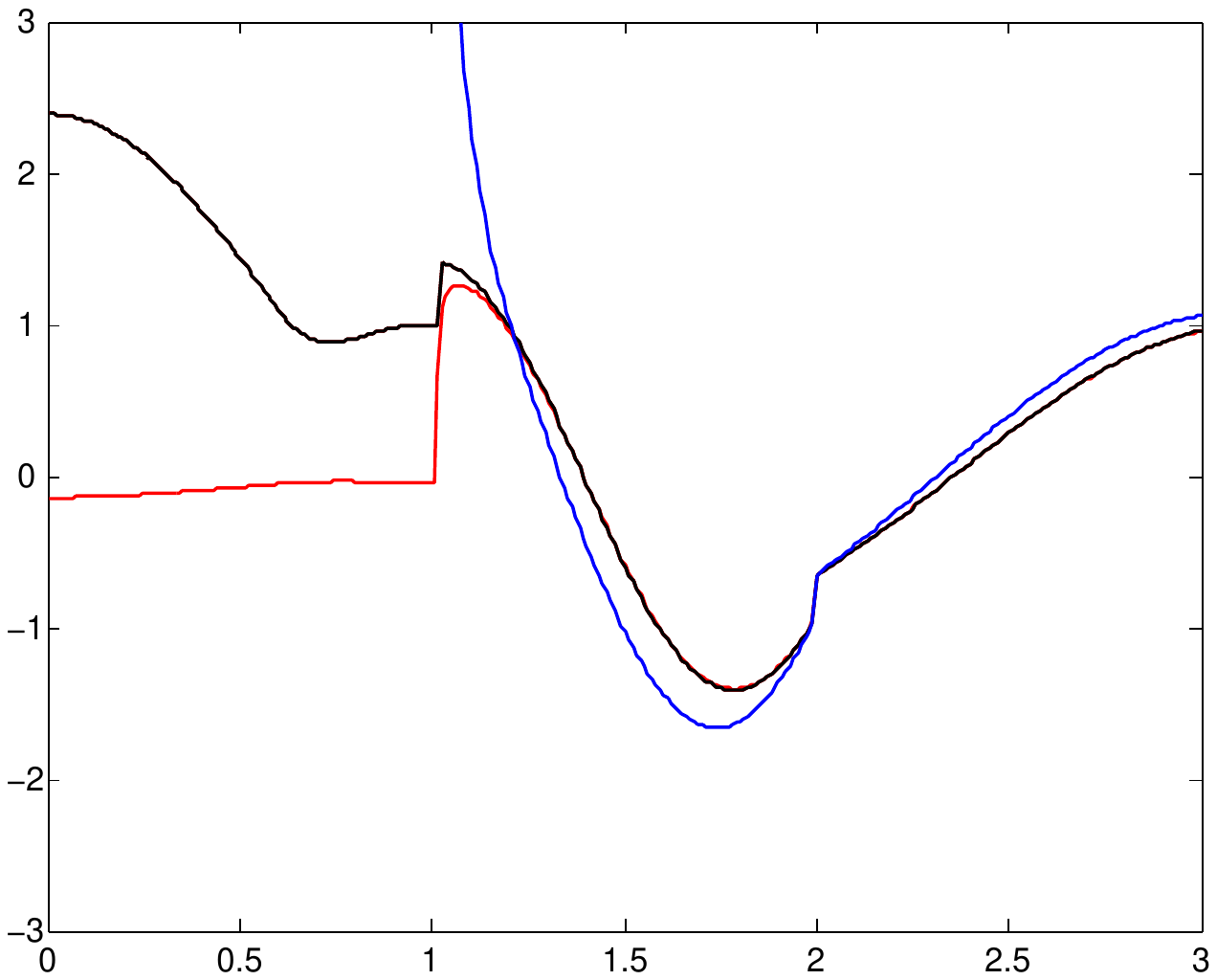}
\hspace{-2.5cm}
 \includegraphics[width=.55\linewidth]{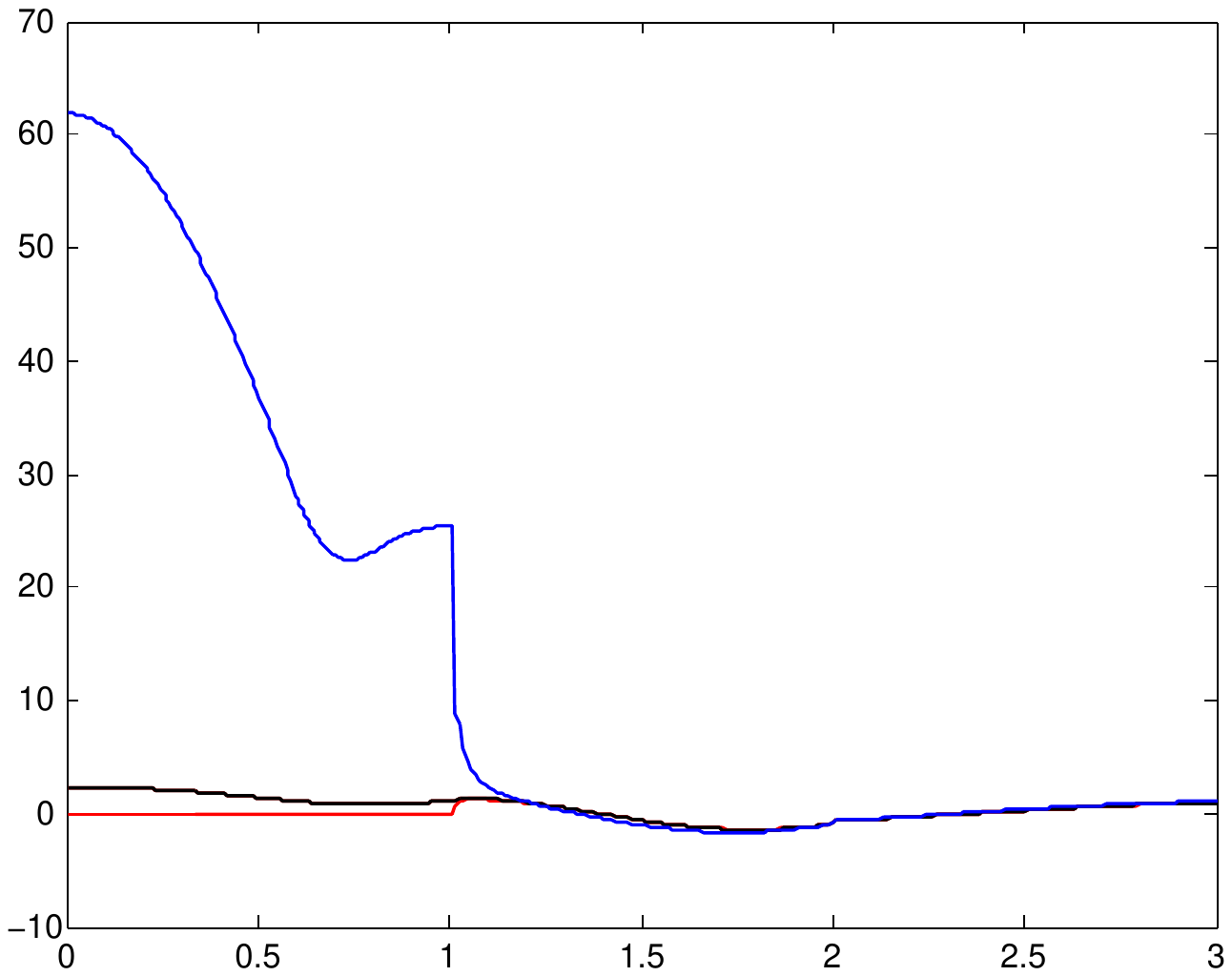}
\end{center}
\vspace{-3.5cm}

\caption{{\bf The cloak - resonance  - Schr\"odinger hat  continuum. } Scattering  by potentials with  different parameters,
demonstrating  three  modes: cloak-, resonance- and Schr\"odinger hat-mode. 
The graphs show the real parts of the effective fields $\psi^{eff}$ on $0<r<3$.
 The same curves are shown on different vertical scales:  {\bf (Left)} in [-3,3], illustrating
 the waves outside the cloak, and
   {\bf (right)} in [-10,70], 
 where the blow up inside the cloak can be seen.
{\bf (Red)} Quantum cloak ({with parameter $\tau_1=\tau_1^{cl}$}), for which  incident wave does not penetrate the cloaked region.
{\bf (Blue)}  Almost trapped wave ($\tau_1=\tau_1^{res}$).  The
cloaking effect is destroyed due to the strong resonance inside the cloak.
{\bf (Black)} Schr\"odinger hat ($\tau_1=\tau_1^{sh}$);  probability mass is almost entirely captured by the cloaked region, yet scattering is negligible.
The  red and black curves are very close to the incident wave on $r>2$, since both a cloak and a SH produce negligible scattering, while a resonant cloak is detectable in the far field.
 The discrepancy shown by the blue curve is due to  the fact that resonances destroy cloaking and can be observed in
the far field; with the parameters here, the difference in $\psi^{eff}$ is small,
but  the destructive effect  on cloaking can be much stronger.
 For parameters used, see discussion  of numerical simulations in the Supplement.
 }
\end{figure}

 \begin{figure}
 \vspace{-3cm}
 \begin{center} \hspace{-3cm}
  \includegraphics[width=.5\linewidth]{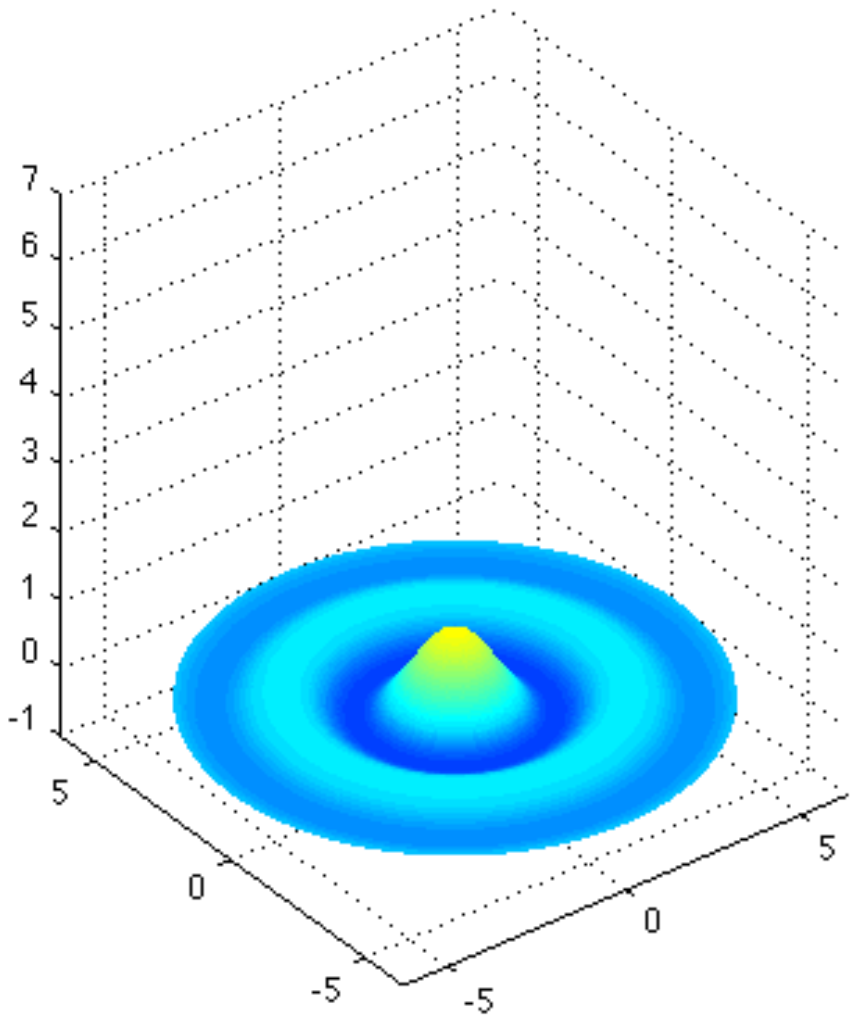} \hspace{-3cm}
  \includegraphics[width=.5\linewidth]{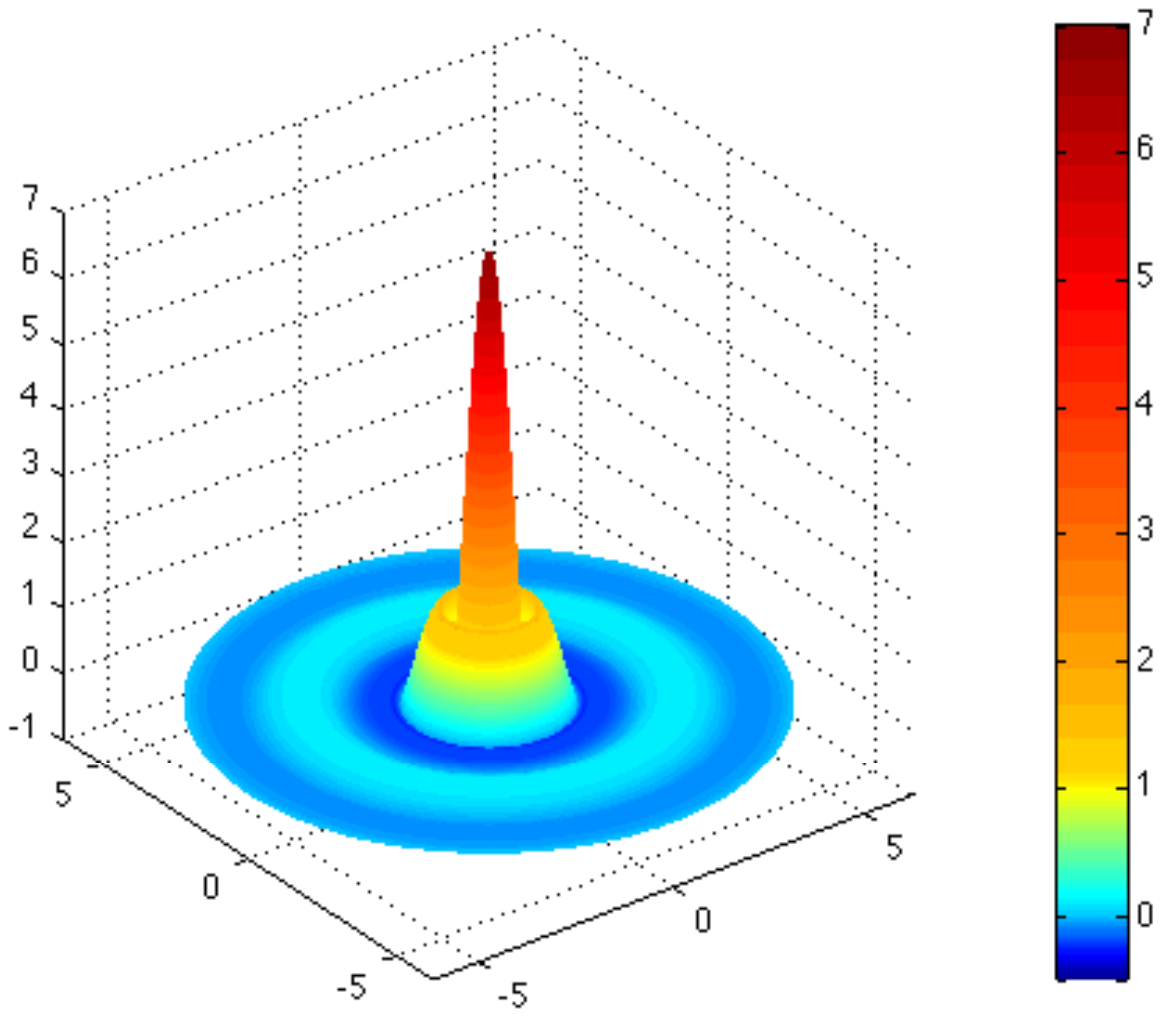} \hspace{-3cm}
  \includegraphics[width=.5\linewidth]{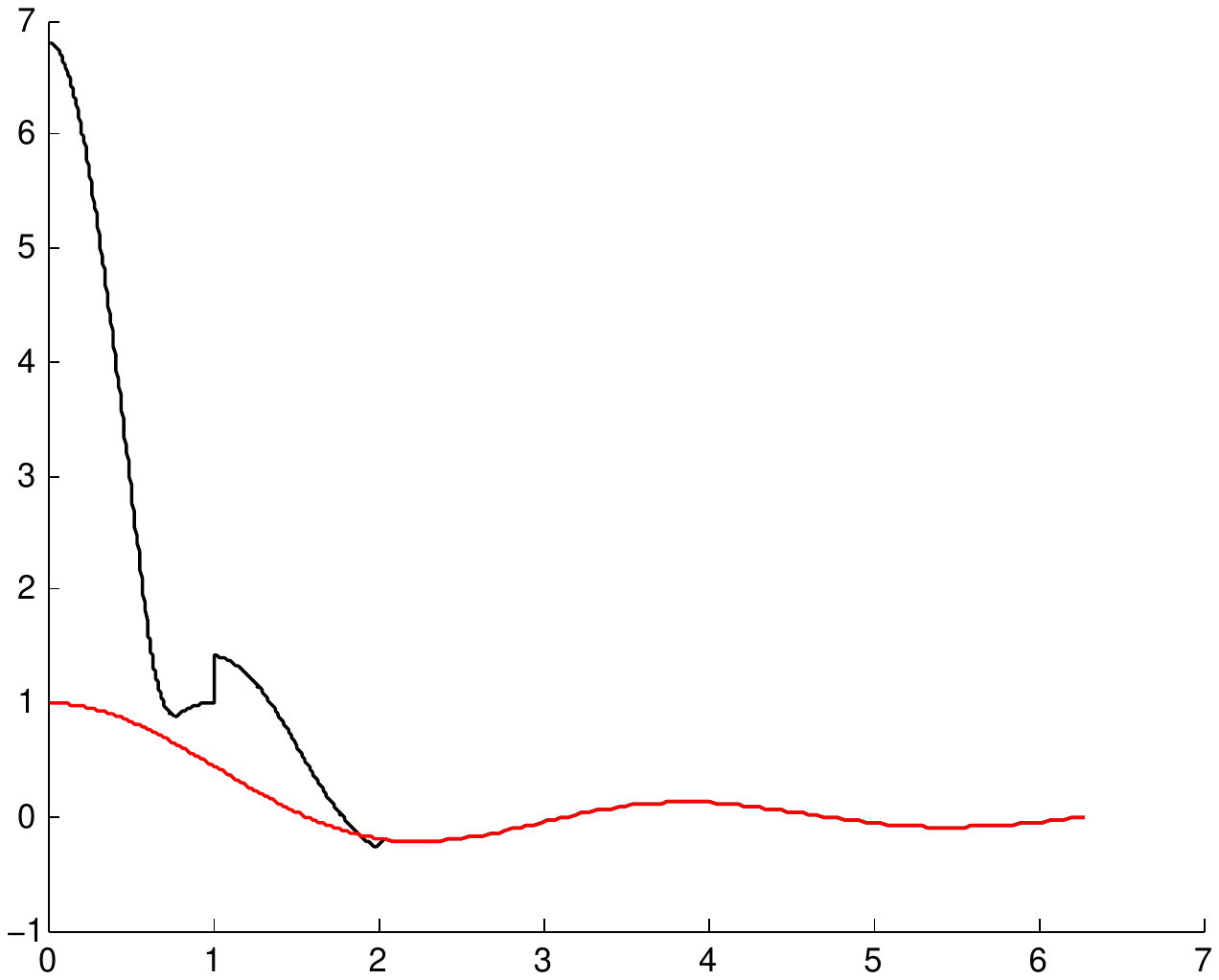}
 \hspace{-10cm}
\end{center}
 \vspace{-3cm}
\caption{{\bf Quantum Three Card Monte.} The 
{ field $\psi^{em}$ ({\bf left}) and  the effective field $\psi^{sh}_{eff}$ ({\bf center})
in the plane $z=0$ when particles are confined in a ball
$B_L$. The central concentration of the wave function changes the conditional probabilities of the particles being found in regions in the vicinity of the cloak.
({\bf Right})  
The non-normalized probability densities $|\psi^{em}|^2$ ({\bf red}) and  $|\psi^{sh}_{eff}|^2$ ({\bf black}) on the positive $x$-axis.}}

\end{figure}

Ideal (perfect) 3D quantum invisibility
cloaks  at fixed energy $E$ are based on the  behavior of solutions to Schr\"odinger equations, 
with specific potentials and  singular, inhomogeneous and anisotropic mass density \cite{Zhang}.
These are mathematically equivalent, via a Liouville gauge transformation,
 to   Helmholtz equations  which also  allow for cloaking in scalar optics \cite{Le,GKLU1} and acoustics \cite{ChenChan,CummerEtAl,GKLUPreprint}. Realizing  a QM cloak would be  challenging, due to the extreme  material parameters   required \cite{Zhang}.
We have previously described approximate  QM cloaks, avoiding extreme and anisotropic parameters but nevertheless acting with arbitrary cloaking effectiveness 
\cite{GKLU3,GKLU_JST}.
If a matter wave is incident to such  a  potential, 
the scattered wave can be made as small as desired.
Analysis  of approximate QM
cloaks revealed a difficulty:  
the wave vanishes inside the cloak  unless
the cloak supports an almost trapped matter wave (or resonance), whose
existence  destroys the cloaking phenomenon and makes the  `cloaked'   region in fact detectable.
However, approximate    cloaks can be 
tuned with a precise choice of  parameters,
{\it close to but not at} resonance; 
the flow of the wave from the exterior into the cloak
and from the cloaked region out into the exterior are balanced, and the cloaking effect
is not destroyed, but rather greatly improved \cite{GKLUSensor}.  We point out similar but surface-plasmon based effects \cite{AE}
and   other subwavelength plasmonics related to sensing \cite{Zul,Fran}.

{An approximate  QM cloak 
can be implemented as follows,
starting from the ideal 3D spherical transformation optics EM  invisibility
cloak \cite{PSS1}. This is based on the  `blowing up a point' coordinate transformation \cite{GLU1,GLU2} $\y\mapsto \x$,
 \beq\label{first transf}
\x:=F(\y)=\y,\hbox{ for } 2<|\y|\le 3;\quad
F(\y)=\left(1+\frac {|\y|}2\right)\frac{\y}{|\y|},\hbox{ for } 0<|\y|\leq 2. 
\eeq
This works equally well in acoustics, forming  a cloak with a 
spherically symmetric singular anisotropic mass density  and 
singular bulk modulus  \cite{ChenChan,CummerEtAl,GKLUPreprint}.
Consider  the case where the anisotropic mass density, 
$M(\x)$, is the identity matrix, and the inverse of the bulk modulus, $\kappa(\x)$, is
$=1$ outside the layer $1<r<2$;   the cloaked region is  the ball $B_1$ of
radius 1 centered at origin. 
For an arbitrary choice of   $R>1$, the ideal cloak is then approximated, 
replacing both the   mass density and bulk modulus by 1  in the shell (or layer)
$1<r<R$. This gives
a  non-singular mass density $M_R$ and non-singular
bulk modulus $\kappa_R^{-1}$, which  approach the ideal cloak parameters as $R\searrow 1$. 
Via homogenization theory, the anisotropic mass density $M_R$ is  approximable
by  \emph{isotropic} mass densities
$m_\e$, 
consisting  of  shells of thickness $\e$ having alternating  large and small densities, 
yielding a family of approximate cloaks \cite{GKLU_JST}.
One then obtains a QM cloak by applying 
the {\mmtext Liouville}-gauge transformation $\psi(\x)=m_{\e}^{-1/2}(\x)u(\x)$, 
so that  the  Helmholtz equation becomes
the  time independent Schr\"odinger equation,  $(-\nabla\cdotp\nabla+V_c-E)\psi=0$, 
where $E=\omega^2$ is the energy and 
$V_c=(1-m_{\e}\kappa_R)E+m_{\e}^{1/2}\nabla\cdotp\nabla(m_{\e}^{-1/2})$
is the cloaking potential  for the energy level $E$.

{For  acoustic  or  EM cloaks constructed using  positive index materials, 
resonances can allow large amounts of energy to be stored inside the `cloaked' region,
but at the price of destroying the cloaking effect \cite{GKLU3,GKLU_JST}.
However, inserting negative index materials within the cloaked region
allows for the cloaked storage of arbitrarily large amounts of energy;
for simplicity, we describe this primarily in the context of QM cloaking,
where the analogous effect is concentration of probability mass. }
When the cloaking potential  is augmented by an internal potential consisting of a series of $N$  shells, alternating positive barriers and negative wells with  appropriately chosen  parameters,   the probability
of the particle being inside the cloaked region can be made as close to 1 as desired.
More precisely,  insert into $B_1$  a piecewise constant potential $Q(\x)$,
{ consisting of two shells,
with values $\tau_1,\ \tau_2$,  in  
$r\leq s_1,\, s_1<r\leq s_2<1$, resp., and zero elsewhere.} 
For suitable parameters  $R,\epsilon, s_j$ and $\tau_j$ of the potential $W$, we obtain, in the Supplement,
a Schr\"odinger  hat  
potential, $V_{SH}=V_c+Q$. Matter waves incident on the SH are  modeled by   Schr\"odinger's equation,
\beq\label{eq: schr eq2}
(-\nabla^2+V_{SH}-E)\psi=0.
 \eeq
{The key feature of  $V_{SH}$
is that the matter waves governed by  (\ref{eq: schr eq2}) 
{can be  made to concentrate
inside the cloaked region as much  as} 
desired, while nevertheless maintaining the cloaking effect, quantified as follows.
Assume that we have two balls   of 
radius $L>2$,   one  ($B_L^{em}$) empty  space and
another ($B_L^{sh}$) containing a Schr\"odinger hat. Let $B_1^\cdot,B_2^\cdot$ denote the central balls of radii 1 and 2, resp.,
for $\cdot=em$ or $sh$,
 and assume that  matter waves $\psi^{em}$ and $\psi^{sh}$ on $B_L^{em}$, $B_L^{sh}$, resp., 
have the same boundary values  on  the sphere of radius L, corresponding to identical incident waves. 
Define the \emph{strength} of the Schr\"odinger hat to be the ratio
\ba
{\frak S}=\frac{1}{ |\psi^{em}(0)|^2} \int_{B_1^{SH}}|\psi^{sh}(\x)|^2\,d\x.
\ea
where $\psi^{em}(\x)$ and $\psi^{sh}(\x)$ are  solutions which coincide in $|\x|>2$.
We show that, by appropriate choice of the design parameters, ${\frak S}$ may be made to
take any prescribed positive value. For large values of ${\frak S}$, the probability mass of $\psi^{sh}$ is almost completely concentrated in the cloaked region. 
{\ttext  For $\e>0$, the wave $\psi^{sh}$ is rapidly oscillating, and
so we also consider the effective wave, $\psi_{eff}^{sh}$,  which is obtained as the limit of $\e\to 0$ (in a suitable weak sense 
discussed in the Supplement).
} This is distinct from the `mirage effect' for standard cloaks, which makes a source within the cloaking layer appear to be in a different position due to the chain rule \cite{ZGNP,GKLU1}.

{We next describe some remarkable properties of Schr\"odinger's hat. To start with, the highly concentrated part of the wave function
which the SH and the incident wave produce inside the cloaked region can be considered as a quasiparticle,
which we call a \emph{quasmon}.
{\ttext A quasmon has  a well-defined   electric charge and variance of momentum  depending on the parameters of $V_{SH}$.
}} 
{{\ttext Secondly, the
amplification  and concentration of a matter wave in the cloaked region can be used to  create  probabilistic illusions.} 
Consider   (non-normalized) wave functions $\psi^{em}$
and  $\psi^{sh}$ which coincide in 
 $B_L-B_2$, i.e., exterior to the cloaking structure.
Then for any region $R_{out}$ in $B_L-B_2$
the conditional probability that the particle
 is observed to be in $R_{out}$,  given 
that it is  observed  in  $B_L-B_2$, is the same for $\psi^{em}$
and  $\psi^{sh}$.  
However, by choosing the parameters of the SH appropriately (see the Supplement),
 the probability that the particle is in the cloaked region $B_1$
can be made as close to 1 as wished.  
{Roughly speaking, the particle  $\psi^{sh}$  is like a trapped ghost of the particle  $\psi^{em}$
in  that it is located in the exterior of the cloaking structure with far
lower probability than  $\psi^{em}$ is, but when $\psi^{sh}$ is observed in $B_L- B_2$,
all} measurements coincide with those of $\psi^{em}$. When a particle is close to} a $SH$, with a large probability it is grasped by the hat and bound into storage  
 within the  cloaked region
as a quasmon.  This is nevertheless consistent with the uncertainty principle: although the particle is spatially localized within $B_1$, the expected value of the magnitude of its momentum is large, due to the  large gradient of $\psi^{sh}$ on a spherical shell about the central peak; cf. Figs. 1 \& 3(center, right).

The Schr\"odinger hat produces vanishingly small changes in 
the matter wave outside of  the cloak, while simultaneously making  the particle
concentrate inside the cloaked region.
Thus, if the matter wave is charged,  it may couple via Coulomb interaction with other particles or measurement devices
external to the cloak.
{\atxt  When an incident field  $\psi^{in}$ is scattered
by the SH,
 the wave field is not  perturbed outside of the
support of the hat potential  $V_{SH}$;  
there are no changes in
scattering measurements. However,
the cloak concentrates the charge 
inside  the cloaked region,  proportional to the square of the modulus of 
the value $\psi^{in}(O)$
which the incident field would have had at the center  $O$  of $B_L^{em}$ in the absence of the SH. 
 Due to the long range nature of the Coulomb potential,  this charge 
causes an electric field which may be strong even far away from
 the SH. If one measures the  
electric field and the result is zero, then 
this indicates that $\psi^{in}(O)=0$;
without disturbing the field, one determines
whether the incident field vanishes at a given point.
A  measuring device within a Schr\"odinger  hat     thus acts as a non-interacting  sensor, 
detecting the nodal curves {or surfaces} on which the incident matter field $\psi^{in}$ vanishes, an effect
analogous to  cloaked acoustic and EM sensors  \cite{AE,GKLUSensor} and near-field scanning optical microscopes \cite{AE2}. }
As described in the Supplement,  a Schr\"odinger hat potential also amplifies
the interaction between two charged particles.

The behavior of Schr\"odinger hats and quasmons can be  illustrated by means of   a quantum variant of  Three Card Monte,
the classic game of chance in which  a coin is hidden under one of 
three bowls and the player guesses where the coin is.
Consider first  a preliminary  version of the game, played by 
  Alice, who  runs the game, and Bob, who makes the guesses. 
In place  of bowls, they play the game   using $N$ empty balls,
each  a copy of $B_L^{em}$,
and the coin   is replaced by a QM particle.
The surface of each ball is made of material representing an infinite potential wall, so that a
particle within cannot escape; this corresponds to
the Dirichlet boundary condition on the boundary. In the game, 
Alice inserts one particle into one of the balls, after which she mixes the balls randomly
and asks  Bob to guess in which ball the particle is.  Bob  
chooses one  and makes internal measurements  near the
boundary of $B_L$. 
Bob, wishing to determine whether the particle is in a region $R$, $R\subseteq B_L -B_2$, measures 
the value of an observable $A$, which  is 1 if 
the particle is observed  ($X\in R$) and 0 otherwise;
the value
of $A$ is also 0 if the particle is not  in the chosen ball.
The  expected value 
of $A$ 
 is  $ p\,\cdotp \mu^{em}_A+(1-p)\,\cdotp 0=p\,\cdotp \mu^{em}_A$. Here, $p=1/N$ is the probability
that Bob chose the ball that  into which Alice  inserted the  {\mltext particle}, and 
$\mu^{em}_A=a_{em}/c_{em}$,  where $a_{em}=\int_{R} |\psi^{em}(\x)|^2d\x$,
$c_{em}=\|\psi^{em}\|^2_{L^2(B_L)}$,  and $\psi^{em}$  is the empty space  wave function on $B_L^{em}$.

To make the game more interesting,
Alice and Bob make a wager: 
they agree that Bob  will pay $ p\mu^{em}_A$ \euro \,   to Alice in advance of each turn, but if he then observes a particle
will receive 1 \euro   \, back from Alice. With these rules, the game is  fair,
with expected profit   0 for  both Alice and Bob.

{Now suppose that, before play commences, and unbeknownst to Alice and Bob,  a third player (the Cloaker) replaces each of the $N$  empty balls with a ball $B_L^{sh}$ equipped with a Schr\"odinger hat.}
The expectation of  $A$
is now  $p\,\cdotp \mu^{sh}_A+(1-p)\,\cdotp 0=p\,\cdotp \mu^{sh}_A$, where
$\mu^{sh}_A=a_{sh}/c_{sh}$,  where $a_{sh}=\int_{R} |\psi^{sh}(\x)|^2d\x$,
$c_{sh}=\|\psi^{sh}\|^2_{L^2(B_L)}$,  and $\psi^{sh}$  is the   wave function  on $B_L^{sh}$.

{
Since a Schr\"odinger hat is an effective cloak,
 $\psi^{em}=\psi^{sh}$ outside  of the ball $B_2$ which contains $R$,
and so  $a_{em}=a_{sh}$. 
 On the other hand,  the presence of the Schr\"odinger hat amplies the wave function 
in $B_1^{sh}$ and} so  $c_{sh}>>c_{em}$;  hence $\mu_A^{sh}<<\mu_A^{em}$, cf. Fig. 3.
When the game is played many times, Bob's expected chance of observing
the particle  in the ball which he chose is smaller than
it was before the Schr\"odinger hats  were inserted. 
In other words, after the  Cloaker's
intervention  the particles start to disappear from 
Bob's observations and    Bob  starts to lose;   Alice is unknowingly  `cheating'.  
The game can be made as unfair as one wishes
 by choosing parameters
so  that ${\frak S}$ is very large,
using general Schr\"odinger hat potentials as described in the Supplement.

\medskip

We conclude by describing one possible path,
discussed in more detail at the end of the paper, towards a solid state realization of a quantum 
Schr\"odinger hat, utilizing a sufficiently large heterostructure 
 of semiconducting materials.
By homogenization theory, the SH potential   can be approximated 
using layered {potential} well shells of depth $-V_-$ and wall shells of  height $V_+$. By rescaling the 
${\bf x}$ coordinate we can make the values
$V_\pm$ smaller (note that in such scaling the size of the support of the SH potential
grows and $E$ becomes smaller).
This sequence of spherical {potential} walls and wells can be implemented using
a heterostructure of semiconducting materials.
In such a structure the wave functions of electrons
with  energy close to the bottom of the conduction bands
can be approximated using Bastard's envelope function method \cite{Bastard}.
Choosing the materials and  thickness of the spherical layers  suitably, 
the envelope functions then satisfy a Schr\"odinger equation whose solutions
are close to those corresponding to the SH potential. 
}

\bigskip

\newpage


%
%
%
%

\vfil\eject

\centerline{\LARGE\bf Supplemental Material}
\vspace{1cm}

{\bf In this supplement, we provide the rigorous analysis needed to confirm the existence and  behavior of Schr\"odinger hats, specify the parameters used in the Figures, and detail the proposed solid-state implementation.}

\bigskip

 \section{Analysis of Quantum cloaks and Schr\"odinger hat potentials}

\subsection{An approximate acoustic  cloak}

Below we will use the approximate cloaks modeled by the Helmholtz (or the Schr\"odinger) type
equation
\beq\label{equat 1 PRE}
& &\left(-\nabla \cdotp M_R^{-1} \nabla -\omega^2 \kappa_R \right)u_R
=0\quad\hbox{in  the  domain }\Omega\subset \R^3,\\
& &u_{R}=h\quad\hbox{on the boundary  $\p \Omega$},\nonumber
\eeq
where $R>1$ is a parameter corresponding to the effectiveness of the cloak, $\omega$ is the frequency, and 
$M_R$ and $\kappa_R$ are the coefficient functions defined below. Let $\nu$ denote the outward unit normal
vector of $\p \Omega$.
Measurements on the boundary $\p\Omega$ are mathematically
modeled by the Dirichelet-to-Neumann operator defined by
\ba
\Lambda_R(h)=\nu\,\cdotp M_R^{-1} \nabla u_R|_{\p \Omega},
\ea
describing the response of the system, i.e., the Neumann boundary
value $\nu\,\cdotp M_R^{-1} \nabla u_R|_{\p \Omega}$, when the Dirichlet data $u_{R}=h$
is posed on the boundary. In the theory of the approximate cloaks the coefficient
functions $M_R$ and $\kappa_R$ are constructed in such a way that as $R\searrow 1$
 the Dirichlet-to-Neumann
operators $\Lambda_R$ approach to  the Dirichlet-to-Neumann operator $\Lambda^{homog}:h\mapsto 
\nu\,\cdotp \nabla u|_{\p \Omega}$ for the boundary value problem
\beq\label{homog. PRE}
& &\left(-\nabla \cdotp  \nabla -\omega^2  \right)u
=0\quad\hbox{in  the  domain }\Omega\subset \R^3,\\
& &u_{R}=h\quad\hbox{on the boundary  $\p \Omega$},\nonumber
\eeq
modeling empty space.
In practical terms, this means that when the parameter $R$ is close to 1,  
for the approximative cloak 
all boundary observations
on $\p \Omega$ are close to the observations on $\p \Omega$ made when
 the domain $\Omega$  is filled with a homogeneous, isotropic medium. 

Approximate cloaks are the basis of our construction
of  Schr\"odinger hat potentials. We start by recalling some facts concerning nonsingular approximations to   ideal  3D spherical cloaks 
\cite{KSVW,GKLU3,GKLU_JST,GKLU_NJP,KOVW,RYNQ}.
For $R>0$, let $B_R=\{|\x|< R\}$ and $S_R=\{|\x|=R\}$ be the open ball and sphere, resp.,  centered at the origin $\mathcal O$ and  of radius $R$ in three-space. Moreover, let $\overline B_R=\{|\x|\leq R\}$ be the closed ball.
For $1\le R<2$,  set $\rho=2(R-1),\, 0\le\rho<2$,  so that $R\searrow 1$ as $\rho\searrow 0$, and introduce
 the coordinate transformation 
 $F_R:B_L- B_\rho\to B_L- B_R$, 
 \beq\label{transf}
\x:=F_R(\y)=\left\{\begin{array}{cl} \y,&\hbox{for } 2< |\y|<L,\\
\left(1+\frac {|\y|}2\right)\frac{\y}{|\y|},&\hbox{for }\rho<|\y|\leq 2. 
\end{array}\right.  
\eeq  
For $R=1$ ($\rho=0$),  this is the singular transformation of \cite{GLU1,GLU2,PSS1}, leading to the ideal transformation optics cloak, while for $R>1$ ($\rho>0$),   $F_R$ is nonsingular  and  leads to a class of  approximate cloaks \cite{RYNQ,GKLU3,GKLU_JST,KSVW,KOVW}. 
Thus,  
if $M_0\equiv\delta_{jk}$ denotes the homogeneous, isotropic mass tensor
tensor,  then, for 
$R=1$, the transformed tensor
becomes an anisotropic singular mass tensor,  $M_{1, jk}(\x)$, on $1 < |\x| <L$, {defined in terms of its inverse,}
\beq\label{eqn-transf law-1}
\left(M_1^{-1}\right)^{jk}(\x)=
((F_1)_*\delta)^{jk}(\y):=\left.
\frac 1{\det [\frac {\p F_1}{\p x}(\x)]}
\sum_{p,q=1}^3 \frac {\p (F_1)^j}{\p x^p}(\x)
\,\frac {\p (F_1)^k}{\p x^q}(\x)  \delta^{pq}(\x)\right|_{\x=F_1^{-1}(\y)}.
\eeq
{\ttext This means that in the Cartesian coordinates $M_1(\x)$ is the matrix with elements
\ba
(M_1)_{jk}(\x)=\frac 12(\delta_{jk}-P_{jk}(\x))+\frac 12(|\x|-1)^{-2}P_{jk}(\x),  \quad 1<|\x|<2,
\ea
where the matrix $P(\x)$, having elements
$P_{jk}(x)=|\x|^{-2}x_jx_k$, is the projection to the radial direction.}

On the other hand, when $R>1$, we obtain an anisotropic but 
nonsingular mass tensor,  $M_{R, jk}(\x)$, on $R < |\x| <L$, given by
\beq\label{eqn-transf law}
\left(M_R^{-1}\right)^{jk}(\x)=
((F_R)_*\delta)^{jk}(\y):=\left.
\frac 1{\det [\frac {\p F_R}{\p x}(\x)]}
\sum_{p,q=1}^3 \frac {\p (F_R)^j}{\p x^p}(\x)
\,\frac {\p (F_R)^k}{\p x^q}(\x)  \delta^{pq}(\x)\right|_{\x=F_R^{-1}(\y)}.
\eeq
For each $R>1$, the eigenvalues of $M_R$ are  bounded from above and below; however, two of them  $\nearrow\infty$
as $R \searrow 1$.
We  define an   approximate  mass tensor $M_R$ everywhere on $B_L$ by {\ttext extending it
as an identity matrix,
\beq \label{R-ideal}
M^{ext}_{R, jk}(\x)=\left\{\begin{array}{cl }M_{R, jk}(\x)
&\hbox{for }  R<|\x|\le 2,\\
\delta_{jk},&\hbox{for }|\x|<R\hbox{ or } 2<|\x|\le L. \end{array}\right.  
\eeq
In sequel, we use the notation $M_{R, jk}(\x)$ also for $M^{ext}_{R, jk}(\x)$.} 
{We  define a scalar function $\kappa_R(\x)$  on $B_L$,
\beq \label{R-equation}
\kappa_R(\x)=
\left\{\begin{array}{cl} 
\eta(\x) ,&\hbox{for }|\x|\leq R_0,\\ 
64|\x|^{-4}(|\x|-1)^4&\hbox{for } R<|\x|<2,\\
1,&\hbox{for }R_0\le |\x|\le R\hbox{ or } 2\le |\x|\le L. \end{array}\right.  
\eeq
where} 
\beq\label{g-alternative}
\eta(\x)=\eta(\x;\tau)=\sum_{j=1}^{N} \tau_j\chi_{(s_{j-1},s_{j})}(|\x|).
\eeq
Here $\tau=(\tau_1,\tau_2,\dots,\tau_N)$, $\tau_j\in \R$ are parameters which one can vary,
$0=s_0<s_j<s_{N}=R_0 $ are some fixed numbers,
and $\chi_{(s_{j-1},s_{j})}(r)$ is the indicator function of
the interval $(s_{j-1},s_{j})$. This means that we have a homogeneous ball $B_{s_0}$
coated with homogeneous
shells.
{\mmmtext Sometimes we denote $\kappa_R(\x)=\kappa_R(\x;\tau)$.} 
Note that   in acoustics $\kappa_R$ has the meaning of inverse of bulk modulus; later,  in quantum mechanics, it  gives rise to the potential.

{\mtext
Below, we  consider what happens as $R\searrow 1$.
In fact, for rigorous mathematical analysis 
we should modify the above definition of
$\kappa_R$ by replacing it, e.g.,\  by $\kappa_1$
so that for all $x\in B_L$
 the quadratic form corresponding
to operator $\int_{B_L}(\nabla v\,\cdotp M_R^{-1} \nabla \overline v-\omega^2 \kappa_R|v|^2)d\x$
becomes smaller (for any fixed $v$) as $R$ decreases.
However, in order to compute solutions explicitly and to present
considerations in a simplified way, we will consider 
 the case when $\kappa_R$ is defined as above. 
 {Mathematical proofs 
will be presented elsewhere.}

Next,  consider in the domain $B_L$ the solutions of the  Dirichlet problem,
\beq\label{equat 1}
\left(-\nabla \cdotp M_R^{-1} \nabla -\omega^2 \kappa_R \right)u_R
=0\quad\hbox{in }B_L,\quad
u_{R}|_{S_L}=h.
\eeq
Since, {\mltext for $1<R<2$, the matrix $M_R$ is  nonsingular everywhere, 
across the internal interface  $S_R$ we have 
the standard transmission conditions, 
\beq\label{trans a1}
& & u_R|_{S_{\radius^+}}=u_R|_{S_{\radius^-}},\\  \nonumber
& &  
{\bf e_r}\cdotp \left(M_R^{-1} \nabla u_R \right)|_{S_{\radius^+}}=
 {\bf e_r}\cdotp \left(M_R^{-1} \nabla u_R \right)|_{S_{\radius^-}},
\eeq
where ${\bf e_r}$ is the radial unit vector and $\pm$ indicates  the 
trace on $S_R$ as $r\to R^\pm$.

In the physical space $B_L$  one has
 \beq \label{24.9.5}
u_R(\x)=\left\{\begin{array}{cl} v_R^+\left(F_R^{-1}(\x)\right),&\hbox{for } R<|\x|<L,\\
 v_R^-(\x),&\hbox{for } |\x|\leq R,\end{array}\right.  
\eeq
with $v_R^{\pm}$ in the virtual space, which consists of the disjoint union $(B_L-B_{\rho})\cup B_{R}$, satisfying 
\ba
(-\nabla^2-\omega^2)v_R^+(\y)&=&0 \quad \hbox{for
}\rho<|\y|<L,\\ v_R^+|_{S_L}&=&h,
\ea
and
\beq \label{extra-equation}
& &(-\nabla^2-\omega^2\kappa_R)v_R^-(\y)=0, \quad \hbox{for }|\y|<R.
\eeq
With respect to   spherical coordinates $(r,\theta,\varphi)$, 
 the transmission conditions (\ref{trans a1})  become
\beq \label{trans a2}
& &v_R^+(\rho,\theta, \phi)=v_R^-(R,\theta, \phi),
\\ \nonumber
& &\rho^2\, \p_rv_R^+(\rho,\theta, \phi)= R^2 \, \p_rv_R^-(R,\theta, \phi).
\eeq
Since $M_R,\, \kappa_R$ are spherically symmetric, cf. (\ref{R-ideal},\ref{R-equation}),
we can separate variables in (\ref{equat 1}), representing $u_R$ as
\beq \label{24.9.10}
u_R(r, \theta, \phi)= \sum_{n=0}^\infty\sum_{m=-n}^n u_{R}^{nm}(r) Y_n^m(\theta, \phi),
\eeq
where $Y^m_n$ are
 the standard spherical harmonics. Then equations (\ref{equat 1}) give rise to a family
 of boundary value problems for the $u_R^{n m}$.  For our purposes, the most important one is 
  the lowest harmonic term (the $s$-mode), $u_R^{0, 0}$, i.e.,  the radial component of $u_R$, which is  independent of $(\theta, \phi)$. This  is studied in the next section.
 
  \subsection{Spherical harmonic coefficients}
  
 {\bf The lowest harmonics.}
For $R_0<1<R<2$,  consider the Dirichlet problem on the ball
$B_L $,
\beq\label{equat A1}
& &(-\nabla \cdotp  M_R^{-1} \nabla -\omega^2 \kappa_{R})u_{R}
=0\quad\hbox{in }B_L,\\ \nonumber
& &u_{R}|_{S_L}=h(\x).
\eeq
We will express asymptotics in terms of the quantity $\rho=2(R-1)$ as $\r\searrow 0$. 

We have  shown elsewhere \cite{GKLU_JST} that
\begin{itemize}
\item
For a specific value of the parameter $\tau\in \R^N$, denoted $\tau=\tau^{res}(R,\omega)$,
 there is a blow-up effect, or interior resonance, destroying cloaking.
 This corresponds to the case when $\tau$ is such that there 
 equation (\ref{equat A1}) has a non-zero radial solution with $h=0$.
 In this case  the solution $u_R$ grows very much inside the cloaked region
as $R\to 1$. {his means that the inside of the cloak is in resonance and {\ntekst
the wave tunnels outwards through the cloak, so that}
 this resonance is detected by  boundary
measurements outside of  the cloak.}
 
\item
For another specific value of $\tau$, denoted $\tau=\tau^{sen}(R,\omega)$,
the cloak acts as an approximate cloak and inside the cloaked
region the solution is proportional  to the value which the field in 
the empty space would have at the origin. This corresponds to the 
case when the equation
(\ref{equat A1}) has a radial   solution $u_{R}$ which satisfies 
$\p_r u_{R}(L)= \omega j_0'( \omega L)$ and $u_{R}(L)=j_0(\omega L )$, or equivalently,
$u_{R}(\x)=j_0( \omega |\x| )$ for $R<|\x|<L$. 

\end{itemize}
Due to the transmission condition (\ref{trans a1}) we see that
the values $\tau^{res}(R,\omega)$ and $\tau^{sen}(R,\omega)$
are close and $\lim_{R\to 1}\tau^{sen}(R,\omega)=\lim_{R\to 1}\tau^{res}(R,\omega)$.


 We now explain in detail how to choose $\tau=\tau^{sen} (R,\omega)$:
First, fix $R$ and $\omega$, and choose $0<R_0<1$.}
Consider the  ordinary differential equation corresponding
to the radial solutions $u(r)$ of the equation (\ref{equat A1}),
that is,
\beq\label{ODE}
-\frac 1{r^2}\frac d{ dr}\bigg (r^2\sigma_R(r) \frac d{ dr}u(r)\bigg)-\omega^2\kappa_{R}(r)u(r)=0,
\eeq
and pose the Cauchy data (i.e. initial data) at $r=L$,
$u(L)=j_0(L\omega)$, $\p_r u(L)=\omega j_0'(L\omega)$.
{\mmmtext Here, $\sigma_R(r)$ is the $rr$-component of the matrix $M^{-1}_R$,
that is, $\sigma_R(r)=2(r-1)^2$, for $R<r<2$ and $\sigma_R(r)=1$, elsewhere.}
 Then we
solve the initial  value problem for the 
the  ordinary differential equation (\ref{ODE})
on interval $r\in [R_0,L]$  and find the Cauchy data $(u(R_0), \frac {du}{ dr}(R_0))$ at $r=R_0$. Note
that on the interval $r\in [R_0,L]$, $\kappa_R$ does not depend on $\tau$. Consider
next the case when 
\beq\label{choosing tau}
 N=2,\quad s_1=R_0/2,\quad s_2=R_0,\quad\tau=(\tau_1,\tau_2), 
\eeq
where $\tau_1$ and $\tau_2$ are constructed in the following way:

\noindent
First, we choose $\tau_2$ to be a negative number with a large absolute 
value. We then solve
of the initial value problem for (\ref{ODE}) on interval $r\in [s_1,R_0]$ 
with initial data 
$(u(R_0), \frac {du}{ dr}(R_0))$ at $r=R_0$.
In particular, this determines the Cauchy data
$(u(s_1), \frac {du}{ dr}(s_1))$ at $r=s_1$.

\noindent
Secondly, consider  $\tau_1, \tau_2$, as well as $R,R_0$,  to be  parameters, and solve
the initial value problem for  (\ref{ODE}) on interval $r\in [0,s_1]$ 
with initial data 
$(u(s_1) ,\frac {du}{ dr}(s_1))$ at $r=s_1$. Denote the solution
by $u(r;\tau_1,R,R_0,\tau_2)$ and find
the value $\frac {du}{ dr}(r;\tau_1,R,R_0,\tau_2)|_{r=0}$.
Then, for given $R,R_0$, and $\tau_2$, we find $\tau_1>0$ satisfying
 \beq\label{eq: tau_1 equation}
 \frac {du}{ dr}(0;\tau_1,R,R_0,\tau_2)=0.
 \eeq
We choose $\tau_1$ to be the smallest value for which  (\ref{eq: tau_1 equation})  holds, and
 denote this solution by $\tau_1(R,R_0,\tau_2)$.
  Summarizing the above computations, we have obtained a
  cloak  at the frequency $\omega$, that is, for the energy $E_0=\omega^2$, such that
  its radial solution $u(\x)$ satisfies
  $u(\x)=j_0(\omega |\x|)$ for $|\x|\in [2,L]$. Moreover, when $\tau_2$
  is large,  this solution $u_R(r)$ grows {\ntekst exponentially} fast on
  the interval $[s_1,s_2]$, as $r$ becomes smaller, while
  on the interval $[0,s_1]$ it satisfies $\frac {du_R}{ dr}(0)=0$,
 {so that $u(r)$ defines a {\ntekst smooth} spherically symmetric solution of
(\ref{equat A1}). }
  
 In the context of QM cloaks below,
  the construction above  can be considered as follows:
   {\ttext Inside the cloak there is a potential
  well of the depth $\tau_1$, enclosed by a potential wall having the height $\tau_2$. 
  The parameters $\tau_1$ and $\tau_2$ are chosen so that
  the solution is large inside the cloak due to the resonance there. Moreover, 
    the choice of the parameters
 is such that the  {\ttext flow} associated to the wave function from the outside
  into the cloaked region and from the cloaked region to the outside are in balance.}
The cloaked region is thus  well-hidden even though the solution
  may be very large inside the cloaked region. {In a scattering
  experiment,  with high probability the 
  potential captures the incoming particle, but due to the  chosen
  parameters of the
  cloak, external measurements cannot detect this.}

{\mmmtext
Using  the implicit function theorem, one can show  that} 
for generic values of $\tau_2$ and $R_0$, there is a limit
$
\lim_{R\searrow 1} \tau_1(R,R_0,\tau_2)=\tau_1(R_0,\tau_2).
$
We note that  the solution $u_R(r)=
u(r;\tau_1(R,R_0,\tau_2),R,\tau_2)$ of (\ref{ODE})
has limit  $
\lim_{R\searrow 1}u(r;\tau_1(R,R_0,\tau_2),R,\tau_2))
= c\,\Phi(r),\quad r<1$,
{\ttext where} $c\in \C$ and $\Phi(r)\not\equiv 0$  an eigenfunction
of the boundary value  problem
 \beq \label{extra-equation BB2}
& &(\nabla^2+\omega^2\kappa_1(\y;\tau))\Phi(\y)=0, \quad \hbox{for }|\y|<1,
\nonumber\\
& &\p_r \Phi(\y)|_{r=1}=0,
\eeq
{\mmmtext where $\tau=(\tau_1(1,R_0,\tau_2),\tau_2)$ and $\Phi$ is normalized 
so that $\|\Phi\|_{L^2(B_1)}=1$.}

{\bf Higher order harmonics.}
Let $\tau=(\tau_1(R_0,\tau_2),\tau_2)$. As  
 $\tau_1$ was chosen
to be the smallest solution of (\ref{eq: tau_1 equation})
we have that $\omega$ an eigenfrequency 
 of the problem
  \beq \label{extra-equation BB aux}
& &(-\nabla^2-\omega^2\kappa_1(r;\tau))v(\y)=0, \quad \hbox{for }|\y|<1,
\nonumber\\
& &\p_r v(\y)|_{r=1}=0.
\eeq

Let us now analyze the solution (\ref{24.9.5}) using spherical harmonics.
{\ntekst Recall that in $B_R$ the function $v_R^-(\y)=u_R(\y)$ is a solution to the homogeneous equation
(\ref{extra-equation}). Thus, in particular,  
\beq \label{8.5.1}
v_R^-(r,\theta,\varphi)=
\sum_{n=0}^\infty\sum_{m=-n}^n u_R^{n m}(r) Y^m_n(\theta,\varphi),
\eeq
where
\beq \label{8.5.1a}
& &u_R^{n m}(r)=a_{nm}j_n(\omega r)+p_{nm}h^{(1)}_n(\omega r),\quad R_0<r<R,
\\ \nonumber
& &u_R^{n m}(r)=\widehat a_{nm}j_n(\omega {\sqrt \tau_2} r)+
\widehat p_{nm}h^{(1)}_n(\omega {\sqrt \tau_2} r),\quad R_0/2<r\leq R_0,
\\ \nonumber
& &u_R^{n m}(r)=
{\widetilde a}_{nm}
j_n(\omega  {\sqrt \tau_1}r),\quad 0 <r\leq R_0/2,
\eeq
{\nntext and ${\sqrt \tau_2}$ is pure imaginary.}
Here  ${\widetilde a}_{nm}={\widetilde a}_{nm}(\omega; R),\,
a_{nm}=a_{nm}(\omega; R),\,p_{nm}=p_{nm}(\omega; R)$ etc.\ are yet undefined coefficients.
Note that the terms with $h^{(1)}_n(\omega  {\sqrt \tau_1}\,r)$ are absent near $r=0$
since $v_R^-(\y)$ has no singularity at $0$.}

Now, for  $\rho<r< L$,  
\ba
v^+_R(r,\theta,\varphi)&=&
\sum_{n=0}^\infty\sum_{m=-n}^n (c_{nm}h^{(1)}_n(\omega r)
+b_{nm}j_n(\omega r))Y^m_n(\theta,\varphi), 
\ea
with as yet unspecified $b_{nm}=b_{nm}( \omega; R)$ and
$c_{nm}=c_{nm}(  \omega; R)$.

Expand  the boundary value $h$ on $\p B_L$ in surface spherical harmonics as
\beq \label{8.5.5}
h(\theta,\varphi)
=\sum_{n=0}^\infty\sum_{m=-n}^n f_{nm}Y^m_n(\theta,\varphi).
\eeq
As shown in  the previous section, $b_{00}=f_{00}$, $c_{00}=0$,
and both $p_{00}$ and $a_{00}$ can be solved for  using the transmission conditions,
which determines the coefficients for $n=0$.

Next we consider the higher-order coefficients, for $n\geq 1$.
{\ntekst To simplify notations,
denote by  $U_R^{n m}(r)Y^m_n(\theta,\varphi)$ the solution of (\ref{extra-equation})
for which $U_R^{n m}(r)=j_n(\omega\sqrt {\tau_1}r)$ for $r<R_0/2$.
Then the  coefficients in (\ref{8.5.1})
can be written in the form
$
u_R^{n m}(r) ={\widetilde a}_{nm} U_R^{n m}(r).
$
Observe that there exist the limit
\ba
\lim_{R\to 1}U_R^{n m}(r) =U_1^{n m}(r)
\ea
and due to the way  the coefficient $\tau_1=\tau_1(R_0,\tau_2)$ was chosen, 
for generic values of $\tau_2$ we have 
\beq\label{eq: non-vanishing}
\p_r U_1^{n m}(1) \neq 0\hbox{ for all }n \geq 1,\,
-n \leq m \leq n.
\eeq Next we assume that $\tau_2$ is such that (\ref{eq: non-vanishing}) holds.

By the  transmission condition (\ref{trans a1}),
\beq \label{8.5.3}
& &{\widetilde a}_{nm}(R) U_R^{n m}(R)=b_{n m}(R)j_n(\omega \rho)+c_{n m}(R) h_n^{(1)}(\omega \rho),
\\ \nonumber
& & R^2 {\widetilde a}_{nm}(R) \p_r U_R^{n m}(r)|_{r=R}=\rho^2 
b_{n m}(R)\p_r j_n(\omega r)|_{r=\rho}+c_{n m}(R) \p_r h_n^{(1)}(\r)|_{r=\rho}.
\eeq
Using asymptotics of Bessel and Hankel functions \cite{Abramowitz},
we obtain from (\ref{8.5.3}) that
\beq\label{c-b eq}
& &c_{n m}(R)=\frac{2^n\,n !}{(2n+1)(2n)! } {\widetilde a}_{nm}(R) \p_r U_1^{n m}(1) \omega^{n+1}
\rho^{n}+O(\rho^{n+1)}),\\ \nonumber 
& & b_{n m}(R)=-\frac{(2n)!}{2 n! } {\widetilde a}_{nm}(R) \p_r U_1^{n m}(1) \omega^{-n}
\rho^{-(n+1)}+O(\rho^{-n}).
\eeq
Note that above $\p_r U_1^{n m}(1)$ is non-vanishing by (\ref{eq: non-vanishing}).
Using (\ref{c-b eq}) and the Dirichlet condition (\ref{8.5.5}), we  finally see that
\beq \label{8.5.4}
{\widetilde a}_{nm}(R)= (2n+1) \a_n f_{n m} \omega^n \frac 1{j_n(\omega L)} \rho^{n+1}
+O(\rho^{n+2}).
\eeq
Together with the transmission
conditions on $r=R_0$ and $r=R_0/2$, this implies that
\beq \label{8.5.2}
& &b_{mn}= \frac 1{j_n(L\omega)} f_{mn}+O(\rho), \quad c_{mn}=O(\rho^{2n+1}),\\ 
\nonumber
& &a_{mn},\ \widehat a_{nm},\ \widetilde a_{nm}=O(\rho^{n+1}), \quad p_{mn},\ \widehat p_{nm}=O(\rho^{n+1}).
\eeq

The above considerations for $n=0$ and for $n\geq 1$ can be summarized
as follows: As $R\searrow 1$, the solutions $u^R(\x)$ converge in $B_L - B_2$ to the solution $u$ corresponding to  the 
homogeneous {virtual} space,
\beq\label{equat 1Cauxialiary}
(-\nabla \cdotp \nabla -\omega^2 )u
&=&0\quad\hbox{in }B_L\\ \nonumber
u|_{S_L}&=&h(\x)
\eeq
and, in the domain
$B_1$, to the solution in 
empty space,
\beq\label{equat 1Cauxialiary inside}
\lim_{R\to 1}u^R(\x)=\beta u(0)\Phi(\x)  
\eeq
where $\Phi(\y)=\Phi(r)$ is the radial solution of the equation (\ref{extra-equation BB aux}),
$\beta=\frac 1{\Phi(1)}$ and $u(0)$ is the value of the solution of (\ref{equat 1Cauxialiary})
at the origin.

}
Next we consider the implications of this for quantum mechanics.

\subsection{Approximate isotropic cloaks in quantum mechanics - Schr\"odinger's hat potential}

The approximate anisotropic cloak $(M_R,\kappa_{R})$ can be further approximated
by an isotropic cloak $(m_{R,\e},\kappa_{R})$, where 
$m_{R,\e}(\x)$ is a  smooth isotropic (i.e., scalar-valued) mass density,  which  we denote  by lowercase $m$
to distinguish it  from the  anisotropic mass tensor denoted by $M$. It satisfies $
 C_1(R, \e)>m_{R,\e}(\x) >c_1>0$, and
leads to an approximate cloak equation,
\beq\label{equat 1C}
\left(-\nabla \cdotp \frac{1}{m_{R,\e}} \nabla -\omega^2 \kappa_{R}\right)u_{R,\e}
&=&0\quad\hbox{in }B_L\\ \nonumber
u_{R,\e}|_{S_L}&=&h(\x).
\eeq

{\nntext We will use isotropic mass densities which, for $R<|\x|<2$, are
of the form
\beq \label{20.06.11}
\frac{1}{m_{R,\e}(\x)}=a(|x|) p_1\left(\frac {|\x|-R}\e \right)+
b(|\x|)p_2\left(\frac {|\x|-R}\e \right)+
p_3\left(\frac {|\x|-R}\e \right). 
\eeq
Here, $p_1,\,p_2,\,p_3$ are bounded 
non-negative smooth functions with period one such that
$p_1,\, p_2=0$ 
near integer values, while $p_3=1 $ near integer values. 
For each $\r>0$ we choose a sequence of
$\e_k=\e_k(\r) \to 0$ as $k \to \infty$ {\mmmtext such that 
$
\frac{2-R}{\e_k(\r)} 
$ is an integer.}
As for $|\x|\leq R$ and $|\x|\geq 2$, we take $m_{\r,\e}(\x)=1$. It is possible 
to choose $a(|\x|)=O(1)$ and $b(|\x|)=O(|\x|-1)$ {\mmmmtext as $|x|\searrow 1$},  so that
\begin{itemize}
\item[1.]
The  $m_{\r,\e_k}(\x)$ are smooth functions in $B_L$;
  \item[2.]
The $m_{\r,\e_k}(\x)$
 approximate $M_1$ as $\e_k \to 0$ and then $\r \to 0$. {\mmmtext Namely, the  operators $-\nabla\cdotp m^{-1}_{r,\e}\nabla$
converge {(as described below)}  to 
$-\nabla\cdotp M^{-1}_1 \nabla$ 
as $\e_k\to 0 $ and $\rho\to 0$.
Note that these $a(|\x|),\, b(|\x|)$ are independent of $\r$.}
\end{itemize}

{\mmmtext  Below we use the shorthand notation $\e\to 0$ instead of $\e_k\to 0$.}
Denote
\beq\label{theta function}
& &\tilde \theta(\x)=a(|\x|)\int_0^1p_1(r' )dr'+b(|\x|)\int_0^1 p_2(r' )dr'+
\int_0^1p_3(r' )dr',\quad 1 <|\x|<2,
\\ \nonumber
& &
\tilde \theta(\x)=1, \quad |\x|<1 \,\, \hbox{or}\,\, |\x|>2.
\eeq
}

{\mmmtext We can choose  $p_i(r'),\, i=1,2$ to be smooth functions
that are very close to
the characteristic functions of the intervals $(0, 1/2)$ and $(1/2, 1)$, continued
periodically, while $p_3(r')$ has its support very close to $r'=0$, continued periodically.
To have that the Hamiltonians corresponding to $m_{\r,\e}$ to approximate Hamiltonian
corresponding to $M_1$ (cf.  \cite{GKLU_JST} for analysis of  the $\Gamma$-convergence
and the two-scale convergence of these operators) 
 we need that
\ba
\frac 12 (a(r)+b(r))\approx 2,\quad \frac 12 \left(\frac 1{a(r)}+\frac 1{b(r)}\right)
\approx \frac 1{2(r-1)},
\ea
i.e.,
\beq \label{21.06.11}
a(r) \approx 2 (1+{\sqrt{2-r}}),\,\,b(r) \approx 2 (1-{\sqrt{2-r}}),\quad 1<r<2.
\eeq
Thus, the mass density $m_{\r,\e}(|\x|)$
 corresponds to two materials occupying layers of equal width in each cell with those cells
 separated by a very thin layer of uniform density. }

Making a 
{\mmmtext  Liouville} gauge transformation for equation (\ref{equat A1}),
that is, introducing
\beq\label{basic gauge}
\psi_{\r,\e}(\x)=m_{\r,\e}^{-1/2}(\x)u_{R,\e}(\x),
\eeq
 {(\ref{equat 1C}) becomes the Schr\"odinger equation with  potential
 $V_{\r,\e},\,\hbox{supp}(V_{\r,\e}) \subset \overline B_2 - B_1$.
 {\mmmmtext Here, $\hbox{supp}(V_{\r,\e})$,  the {\it support }of $V_{\r,\e}$,
is the set where $V_{\r,\e}(\x)$ is non-zero.}
{For  generic values of the parameters, these model  approximate 
invisibility cloaks for matter waves  \cite{GKLU3}, with Schr\"odinger equations}
\beq\label{eq: schr eq}
(-\nabla\,\cdotp \nabla+V_{\r,\e}+Q_{\r}-E)\psi_{\r,\e}=0 \quad \hbox{in }B_L,
& & u|_{\p B_L}=h.\eeq
Here $E=\omega^2$, and these {\it cloaking potentials}, $V_{\r,\e}$, are of the form,
\beq\label{Vre}
& &V_{\r,\e}(\x)-E=m_{R,\epsilon}^{1/2}\nabla\,\cdotp \nabla (m_{R,\epsilon}^{-1/2})-
E m_{R,\epsilon} \kappa_{R}, \quad R<|\x| <2, \\ \nonumber
& & V_{\r,\e}(\x)=0, \quad 0<|\x| <R \,\,\hbox{and}\,\, 2<|\x| <3.
\eeq
In addition, inside the cloak, i.e. in $B_1$, there is the {\it cloaked potential} (cf. (\ref{g-alternative}))
\beq\label{Qrho}
Q_{\r}(\x)=-E(\eta(\x;\tau)-1), \quad \hbox{supp}(Q_\r) \subset \overline B_{R_0}\subset B_1,
\eeq
where $\tau=\tau^{sen}(R,\omega)$ and
 $Q_{\r}$ depends on the parameter $\tau(R,\omega)$. Thus,
when $\tau $ is appropriately chosen,
the total potential $V_{\r,\e}+Q_{\r}$
in $B_L$ acts as a quantum mechanical sensor cloak.
{It is this \emph{total} potential  which we call a {\it Schr\"odinger hat} (SH).}
{\ttext (The  cloaked potential, that is, $Q_\r$, resembles a  ``hat", specifically the 
Mexican hat, but the terminology is chosen because
of the {\mltext interesting} effects of the potential, not because of its profile.)}

\subsection{Analysis of Schr\"odinger's hat: convergence of normalization constants}

{\mltextt
{To consider  the properties of a Schr\"odinger's hat, one first  needs
some convergence properties of  approximate cloaks as they approach an ideal one,
initially treating the  case when $E$ is not a Dirichlet eigenvalue
of Laplacian in the ball of radius $L$. The  case of Dirichlet eigenvalues will be considered
later.}

 Let $u$ satisfy the boundary value problem
 \beq\label{Schr 2}
& &(-\nabla\cdotp \nabla-E)u=0,\quad\hbox{in }B_L\subset \R^3,\\
& &\nonumber u|_{\p B_L}=h.
\eeq
Moreover, let 
\ba
Q_0= \lim_{\r\to 0}  Q_{\r}= E(1-\eta(\x;(\tau_1(R_0,\tau_2),\tau_2))),
\ea 
(see (\ref{Qrho})),
be the  potentials supported in $B_{R_0}$, {with the parameters suppressed.} 
{Then the $Q_0$  describe QM
 cloaked   sensor  potentials at energy $E$.}

Let $F=F_R$ with $R=1$ be the singular blow up map and define
\beq \label{33a}
\tilde u(\x)=\left\{\begin{array}{cl} u(F^{-1} (\x)),&\hbox{for } \x\in B_L- \overline B_1,\\
\beta u(0) \Phi(\x),&\hbox{for } \x\in \overline B_1, 
\end{array}\right.
\eeq
{\ttext where $\beta={\Phi(1)}^{-1}$ and
  $u$ is the solution of (\ref{Schr 2}).}

Let $u_{\r}$ be a solution of Helmholtz equation
\beq\label{Helm0}
& &(-\nabla\cdotp M^{-1}_{R}\nabla+Q_\r-E\kappa_{R})u_{\r}=0,\quad\hbox{in }
B_L\subset \R^3,\\
& &\nonumber u_{\r}|_{\p B_L}=h,
\eeq
and
 $u_{\r,\e}$ be a solution of Helmholtz equation,
\beq\label{Helm1}
& &(-\nabla\cdotp m^{-1}_{\r,\e}\nabla+Q_{\r}-E\kappa_R)u_{\r,\e}=0,\quad\hbox{in }B_L\subset \R^3,\\
& &\nonumber u_{\r,\e}|_{\p B_L}=h. 
\eeq
{\nntext Then,
\beq \label{7.06.11}
\lim_{\e \to 0} u_{\r, \e} =u_{\r},
\eeq
{\ttext in the space  $L^2(B_L)$}.
Moreover, letting $K\subset B_L-  B_1$ be arbitrary, using (\ref{theta function}), (\ref{basic gauge}),
one can show that
\beq\label{final convergence3} 
\lim_{\rho\to 0}\lim_{\e\to 0}\int_K|\psi_{\r,\e}(\x)|^2 d\x=
\lim_{\rho\to 0}\lim_{\e\to 0}\int_K m_{\r, \e}(\x)^{-1} |u_{\r, \e}(\x)|^2 d\x=
\int_K \tilde\theta(\x)|\tilde u(\x)|^2 d\x .
\eeq
Proofs  of the claims (\ref{7.06.11}) and  (\ref{final convergence3}) are
omitted 
and will be detailed elsewhere, but the principal ideas are that,
 using the above properties of the spherical harmonics expansions,
\beq\label{final convergence1}
\lim_{\rho\to 0}\lim_{\e\to 0}u_{\r,\e}(\x)=\tilde u(\x)\quad\hbox{for }\, \x\in B_L-\p B_1, 
\eeq
and that, for $\phi\in C(\overline B_L)$, 
\beq\label{phi estimatea}
\lim_{\r\to 0}\lim_{\e\to 0}\int_{K}m^{-1}_{\r,\e}(\x)\phi(\x) \,d\x =\int_{K}\tilde \theta(\x) \phi(\x)\, d\x.
\eeq
}

{\mmmtext  In the following, we use
\beq \label{40a} 
\psi^{eff}(\x)=\tilde\theta(\x)^{1/2} \tilde u(\x)
\eeq
which can be  considered as 
the effective wave function, when one is modeling the location
of the particle  corresponding to the wave function $u_{\r,\e}$
with sufficiently small $\r$ and $\e$.} 
 }

\subsection{Conditional probabilities  for location of particles in a Schr\"odinger hat}

 Consider next the situation of two balls of radius $L$, one  empty and
the other with a Schr\"odinger hat potential, and assume that there is 
one particle in each ball. Let $u (\x)$ be wave function corresponding
to the particle in the empty ball and  $\psi_{\r,\e}$ be the 
wave function corresponding
to the ball with the SH potential. Assume that 
 $u$ and $\psi_{\r,\e}$
have  the same boundary value, $h$, on $\p B_L$.

Consider next the normalization constants
\ba
& & C_3^{empty}= \int_{B_L - B_2} |u (\x)|^2\, d\x,\\
& & C_2^{empty}= \int_{  B_2 - \overline B_1} |u(\x)|^2\, d\x,\\
& & C_1^{empty}= \int_{B_1} |u (\x)|^2\, d\x,\\
& & C_3^{cloak}=\lim_{\rho\to 0}\lim_{\e\to 0} \int_{B_L - B_2} |\psi_{\r,\e}(\x)|^2\, d\x,\\
& & C_2^{cloak}=\lim_{\rho\to 0}\lim_{\e\to 0} \int_{B_2 - \overline B_1} |\psi_{\r,\e}(\x)|^2\, d\x,\\
& & C_1^{cloak}=\lim_{\rho\to 0}\lim_{\e\to 0} \int_{B_1} |\psi_{\r,\e}(\x)|^2\, d\x.
\ea
The probability that a free particle $X$ in the empty ball $B_L$, described by the wave function $u$,  
is in fact located  in the smaller ball $B_1$, is equal to 
\ba
\prob(\{X\in B_1\})=\frac {C_1^{empty}}{C_1^{empty}+C_2^{empty}+C_3^{empty}}.
\ea
Similarly, 
 the probability that a particle $\tilde X_{\r,\e}$ in the ball  $B_L$ with the 
SH  potential $V_{\r,\e}+Q_\r$, described by
the wave function $\psi_{\r, \e}$, to be in $B_1$ satisfies 
\ba
\lim_{\rho\to 0}\lim_{\e\to 0}  \prob(\{\tilde X_{\r,\e}\in B_1\})=
 \frac {C_1^{cloak}}{C_1^{cloak}+C_2^{cloak}+C_3^{cloak}}.
\ea
Similar results hold for the probabilities for the particles  to be in 
$B_2 - B_1$ and $B_L\setminus B_2$.

Using (\ref{final convergence3})--(\ref{phi estimatea}),
 we see that
\ba
C_3^{cloak}&=&C_3^{empty},\\
C_1^{cloak}&=&\frac 1{|\Phi(1)|^2}|u(0)|^2.
\ea	
Recall that $u$ is the solution of (\ref{Schr 2}) and $\Phi$ is the $L^2(B_1)$-normalized
 Neumann eigenfunction inside the cloak, cf.\ (\ref{33a}). 
{\ttext As for $C_2^{cloak}$, by (\ref{final convergence3}) and (\ref{40a}), we have
\beq \label{1.06.11}
C_2^{cloak}=\int_{B_2 \setminus B_1} |\psi^{eff}(\x)|^2 d\x =\int_{B_2 \setminus B_1} \tilde\theta(\x)|{\tilde u}(\x)|^2 d\x =
\int_{B_2} \theta(\y) |u(\y)|^2 d\y,
\eeq
\ba
\theta(\y)= \tilde\theta(\x)\, \left|\frac{\p \x}{\p \y} \right|, \quad \x = F(\y), \,\, |\y|< 2.
\ea }
From the definition of $F(\y)$, cf. (\ref{transf}), one sees that
$
\theta(\y)=O\left(\frac{1}{|\y|^2} \right)
$;
however, due to the boundedness of the norm of $u$ in the space 
$H^1(B_L)$, the integral in (\ref{1.06.11})
is bounded.
An important observation is that, by changing the parameters $\tau$
determining the potentials $Q_{\r}, \, Q_0,$ {\it  one can keep $C_3^{cloak}$ and $C_{2}^{cloak}$ 
unchanged
while $C_{1}^{cloak}$ is  made arbitrarily large.}


Consider  a particle $X$ with
energy $E$ in an empty ball $B_L$ and another particle $\tilde X_{\r, \e}$ {\mmmtext with
energy $E$} in the same box 
but with
the  SH potential $V_{\r,\e}+Q_\r$.  Let us next consider the event 
 that the particle $X$ in the empty ball corresponding to wave function $u$
is located in the set $S\subset B_L$, and denote this
event by $L_S$. Then, 
\ba
\prob(L_S)=c_{empty}\int_S |u(\x)|^2\,d\x,\quad c_{empty}=\frac {1}{C_1^{empty}+C_2^{empty}+C_3^{empty}}. 
\ea
The conditional probability for the event that the particle $X$  is in
$S\subset B_L\setminus B_2$, conditioned  on
the particle being in  $S_0:=B_L\setminus B_2$,
is thus
\beq \label{2.06.11}
\prob(L_S|L_{S_0})=\frac {\int_S |u(\x)|^2\,d\x} {\int_{S_0} |u(\x)|^2\,d\x}.
\eeq

Next,   
for a ball $B_L$ in which there is
an SH potential,  denote the event 
 that a particle $\tilde X_{\r,\e}$ 
 is located in the set $S\subset B_L\setminus B_2$ by 
 $\tilde L_S(\r,\e)$.
Then, 
\ba
& &\prob(\tilde L_S(\r,\e))=c_{cloak}(\r,\e)\int_S |\psi_{\r,\e}(\x)|^2\,d\x,\\
& &
\lim_{\r\to0} \lim_{\e\to 0} c_{cloak}(\r,\e)=\frac {1}{C_1^{cloak}+C_2^{cloak}+C_3^{cloak}}.
\ea
Thus, the conditional probability for the event that the particle $X$  is in
$S\subset B_L\setminus B_2$, conditioned on it being in  $S_0:=B_L\setminus B_2$,
is
\beq \label{3.06.11}
\prob(\tilde L_S(\r,\e)|\tilde L_{S_0}(\r,\e))
=\frac {\int_S |\psi_{\r,\e}(\x)|^2\,d\x} {\int_{S_0} |\psi_{\r,\e}(\x)|^2\,d\x},
\eeq
and
\beq \label{4.06.11}
\lim_{\rho\to 0}\lim_{\e\to 0}
\prob(\tilde L_S(\r,\e)|\tilde L_{S_0}(\r,\e))=\prob(L_S|L_{S_0}).
\eeq

In summary, the above computations have the following consequences  for
the wave functions $u$ and $\psi_{\r,\e}$, corresponding to the the particles
in an  empty ball and a ball with the SH potential, resp.
Recall that the  both field $u$ and $\psi_{\r,\e}$
have the boundary value $h$.
Then, under the condition that we observe a particle in $B_L\setminus B_2$,
the conditional probability that the particle is observed to be in 
a set $S\subset B_L\setminus B_2$,  conditioned on
it being  observed to be in 
a set $B_L\setminus B_2$, is same for both balls.

Consider now the case when $S=B_1$, or, more generally, $S \subset B_{R_0} \subset B_1$.
Clearly, equations (\ref{2.06.11})--(\ref{3.06.11}) remain valid for such $S$. However, 
equation (\ref{4.06.11}) is no longer valid. Moreover, by choosing properly parameters $\tau$,
we can make $\lim_{\rho\to 0}\lim_{\e\to 0}
\prob(\tilde L_S(\r,\e)|\tilde L_{S_0}(\r,\e))$ as large as we want.
To measure this effect we introduce the ratio 
\beq \label{11.06.11}
{\frak S}=
\lim_{\r\to 0}\lim_{\e\to 0}\frac 1{|u(0)|^2} \int_{B_1}|\psi_{\r,\e}(\x)|^2\,d\x=
\frac{1}{ |\Phi(1)|^2},
\eeq
see (\ref{33a}), (\ref{40a}), where $\tilde\theta (\x)=1$ in $B_1$;  we call $\frak S$
the \emph{strength} of the SH potential $Q_0$. 
$\frak S$ depends only on the choice of the parameters 
 $\tau(\r,\omega)=(\tau_1,\tau_2)$, in particular, the parameter $\tau_2$ that 
 determines how rapidly
 the solution grows in the layer $B_{s_2} \setminus B_{s_1}\subset B_{R_0}$. 
Choosing  $\tau(\r,\omega)$ appropriately, we can achieve
 any prescribed positive value of $\frak S$. 
 {\ttext The parameters $(\tau_1,\tau_2)$ of the SH potential 
 do not change the (non-normalized) wave function outside the ball $B_1$ but change radically 
 the wave in it. The wave inside the ball $B_1$ has
 the variance of momentum ${\frak S}|u(0)|^2\int_{B_1}|\nabla \Phi(\x)|^2d\x$ and
 the charge $Q'$ considered later in formula (\ref{Q' formula}). We consider the function inside $B_1$
 as a quasiparticle and introduce a solid state model for in Section \ref{sec: Implementation}.}

We analyzed above the eigenfunction in a ball, and  it was enough 
to analyze only the lowest harmonic. {\mmmmtext One can replace the ball $B_L$
with an arbitrary domain $\Omega\subset \R^3$ containing the ball $B_L$  where the cloaked  SH potential is supported using the following observation: The boundary value problem
\beq\label{Omega SH}
& &(-\nabla \cdotp  \nabla+V_{\r,\e}+Q_{\r}-E)\psi_{\r,\e}=0 
=0\quad\hbox{in  the  domain }\Omega,\\
& &\psi_{\r,\e}=h\quad\hbox{on the boundary  $\p \Omega$},\nonumber
\eeq
  is  equivalent to the problem
\beq\label{Omega SH, hole}
& &(-\nabla \cdotp  \nabla-E)\psi_{\r,\e}=0 
\quad\hbox{in  the  domain }\Omega-B_L,\\
& &\psi_{\r,\e}=h\quad\hbox{on the boundary  $\p \Omega$},\nonumber\\
& &\nu\,\cdotp \nabla \psi_{\r,\e}|_{\p B_L}=\Lambda_{\rho,\e} (\psi_{\r,\e}|_{\p B_L}) \quad\hbox{on the boundary  $\p B_L$},\nonumber
\eeq
where $ \Lambda_{\rho,\e}$ is the Dirichlet-to-Neumann operator
for the equation $(-\nabla \cdotp  \nabla+V_{\r,\e}+Q_{\r}-E)\psi=0$ in the ball $B_L$.
The previous analysis for the 
higher order harmonics shows that the  Dirichlet-to-Neumann operators
 $ \Lambda_{\rho,\e}$
for the ball with the SH potential tend, as $\r$ and $\e$ tend to zero, to the  Dirichlet-to-Neumann operator corresponding 
to an empty ball. Thus,  the above analysis 
of  the behavior of the solutions in the ball $B_L$ can be
readily generalized to an arbitrary domain $\Omega$.}
Potentials  which, for \emph{some} incident wave, produce a scattered wave  which is zero outside a bounded set,
are said to have a \emph{transmission eigenvalue} \cite{CoPaSy}.
We emphasize that the scattered wave caused by the SH potential is approximately
zero for \emph{all} incident fields.

An illustrative example of this phenomenon via quantum three-card monte is described in the body of the paper.

\subsection{Scattering from the hat and  particle storage}

Now consider  the effect of Schr\"odingers hats on scattering experiments in $\R^3$. 
{\nntext In the case of free
space, the wave function $\Psi$ satisfies
 \beq\label{Schr scatt}
(-\nabla\cdotp \nabla-E)\Psi=0,\quad\hbox{in }\R^3,
\eeq
and we can choose
\ba
 \Psi(\x)=\Psi^{in}(\x)= e^{i \omega ({\bf e}, \x)},\quad |{\bf e}|=1 ,\,\,
 \omega^2=E,
\ea
so that $ \Psi^{in}(\x)$ are the plane waves in the direction ${\bf e}$.
Compare them with the wave functions in $\R^3$ corresponding to scattering 
from the SH potential,
\beq\label{Schr scattering C}
& &(-\nabla\cdotp \nabla+V_{\r,\e}+Q_\r-E)\Psi_{\r,\e}=0,\quad\hbox{in }\R^3,\\
& &\nonumber \Psi_{\r,\e}=\Psi^{in}+ \Psi^{sc}_{\r,\e},
\eeq
where $ \Psi^{sc}_{\r,\e}(\x)$  satisfies
Sommerfeld's radiation condition. Note that the SH potential 
 $V_{\r,\e}+Q_\r$ vanishes outside $B_2$.}

{\nntext As scattering data for these problems is equivalent to Dirichlet-to-Neumann operators
on $\p B_L$ \cite{Ber},
we can use previous results to consider
scattering from the SH potential. {\mmmmtext  In particular, we see using
 (\ref{basic gauge}), (\ref{33a}), and
(\ref{7.06.11}), that 
 when $\rho$ and $\e$ are small enough, 
the solution $\Psi_{\r,\e}(\x)$ is  close to $\Psi^{in}(\x)$ outside $B_2$,
close to $m^{-1/2}_{\r, \e}(\x)\Psi^{in}(F^{-1}(\x))$ in $B_2-B_R$, and close to $\Psi^{in}(0)\Phi(\x)$
in the cloaked region.
}

Thus, we see that  
 scattering observations, i.e.,  observables depending on
the far field patterns of the solutions, are almost the same for the empty space
and for the Schr\"odinger hat potential (when $\rho$ and $\e$ are small enough).
However, if we consider the {\mltext conditional} probability of particles in
an empty ball and one with with the SH potential,
i.e., the ratios
\ba
I_{S_1,S_2}
=\frac {\int_{S_1} |\psi(\x)|^2\,d\x} {\int_{S_2} |\psi(\x)|^2\,d\x}
\ea
and
\ba
\tilde I_{S_1,S_2}
=\frac {\int_{S_1} |\psi_{\r,\e}(\x)|^2\,d\x} {\int_{S_2} |\psi_{\r,\e}(\x)|^2\,d\x}.
\ea
we see that one can make   $\tilde I_{S_1,S_2}>> I_{S_1,S_2}$
if $S_1$ contains $B_1$ and $S_2$ does not intersect $B_2$. {The physical interpretation of this is
that, when {\mltext a particle}  scatters   from  a Schr\"odinger hat, there is a high concentration
of probability mass in the cloaked region,
but this can not be detected from
far field observations.}

\subsection{A hat and two ions - an interaction amplifier}

A SH potential also amplifies
the interaction between two charged particles. To see this,
 consider two particles of the same species, which for simplicity we simply refer to as \emph{ions}. 
Suppose that one ion, having
energy $E$, is confined to a domain which also includes a SH potential $V_{SH}$,
centered at point $O$.
Tuning the parameters of the SH potential, 
the probability that the
ion is concentrated in a  ball $B_d$, 
of small radius $d$ and center $O$, can be made as close to 1
as desired. If one now inserts {\it two} ions into the domain which interact both  with the 
potential  $V_{SH}$ and with each other via a Coulomb potential, 
both charged particles can not be concentrated in  
$B_d$. First-order perturbation theory shows \cite{Suppl} 
that the  energy level  of the particles in the presence of a SH behaves
as if the charges of the ions were multiplied by a (large) factor $O(d^{-1/2})$.

{
Let us consider a large ball $B_L$ (we could also consider a box $D_L=[-L,L]^3$ or any
other domain) containing a SH potential and
a single  {particle}   system modeled by
 \beq\label{Schr box1}
(-\nabla\cdotp \nabla+V_{\r,\e}(\x)+Q_\r(\x)-E)\psi_{\r,\e}(\x)=0 \,\,\hbox{in }B_L,
\quad \psi_{\r,\e}|_{\p B_L}=0, \quad \|\psi_{\r,\e}\|_{L^2(B_L)}=1,
\eeq
cf. (\ref{eq: schr eq}).
{\nntext  Here $E>0$ is the Dirichlet eigenvalue of {\mmmmtext the partial differential operator} in
(\ref{Schr 2}) in the empty ball $B_L$,
corresponding to  radially symmetric eigenfunction 
$\psi^0(\x)=j_0(\omega|\x|),\,\omega=\sqrt E,$ so that $j_0(\omega L)=0$, and 
$V_{\r,\e}(\x),\,Q_\r(\x)$ are chosen as {\ttext in the subsection.\ I.C.}
}
In addition to the one-particle equation, we consider a   system of two charged 
particles, modeled by 
\beq\nonumber 
& &(-\nabla_x\cdotp \nabla_x-\nabla_y\cdotp \nabla_y+V_{\r,\e}(\x)+Q_\r(\x)+
V_{\r,\e}(\y)+Q_\r(\y)+\frac a{|\x-\y|}-2E')\Psi(\x,\y)=0,\\
\label{Schr. two particle box}
& &\hspace{4cm}\quad\hbox{in }(\x,\y)\in B_L\times B_L,\\
& &\nonumber\hspace{4cm} \Psi|_{\p (B_L\times B_L)}=0,
\eeq
where the energy $2E'$ is close to $2E$. Here, $a$ is proportional to
the product of charges of the ions. 
When $a$ is small (which we now assume), the solution $\Psi$ is a perturbation of $ \Psi^0$, the product of two 
one-particle solutions,
\beq \label{10.06.11}
\Psi^{0}(\x,\y)=\psi_{\r,\e}(\x)\psi_{\r,\e}(\y).
\eeq
By first-order perturbation theory,  
we can write 
\beq\label{scatt. approx.}
& &E'=E+aE_1+O(a^2),\label{scatt. approx.}\\
\nonumber & &
\Psi(\x,\y)=\Psi^{(0)}_{\r,\e}(\x,\y) +a\Psi^{(1)}_{\r,\e}(\x,\y)+O(a^2),
\quad \Psi^{(1)}_{\r,\e}|_{\p (B _L\times B_L)}=0.
\eeq
 We will show that the two-particle interaction is strongly effected by 
 the presence of  a SH.

Substituting  approximation (\ref{scatt. approx.}) in equation (\ref{Schr. two particle box}) and considering
terms of order $O(a)$, we obtain the equation 
\beq\label{2-particle approx}
& &\big(-\nabla_x\cdotp \nabla_x-\nabla_y\cdotp \nabla_y+V_{\r,\e}(\x)+Q_\r(\x)+
V_{\r,\e}(\y)+Q_\r(\y)-2E \big) \Psi^{(1)}_{\r,\e}(\x,\y)\\ \nonumber
& &=
\left(2E_1-\frac 1{|\x-\y|}\right) \Psi^{(0)}_{\r,\e}(\x,\y),
\quad\hbox{for }(\x, \y)\in B_L\times B_L,\,\,  \Psi^{(1)}_{\r,\e}|_{\p (B_L\times B_L)}=0
\eeq
{\nntext As $2E$ is the eigenvalue of the left hand side of (\ref{2-particle approx}) with 
$\Psi^{(0)}_{\r,\e}$ being its eigenfunction, it is necessary that
}
\beq\label{derivative of energy}
& & E_1=\frac 12 \int_{B_L}  
\int_{B_L} \frac 1{|\x-\y|}|\psi_{\r,\e}(\x)|^2
 |\psi_{\r,\e}(\y)|^2\, d\x d\y,
 \eeq
 where we have  used  (\ref{10.06.11}), (\ref{Schr box1}).
 {\mmmtext Let
\ba
\Phi_L(\x)=\int_{B_L}\Psi_{\r,\e}(\x,\y)\overline {\psi_{\r,\e,}(\y)}d\y,
\quad \Phi_L^{(1)}(\x)=\int_{B_L}\Psi^{(1)}_{\r,\e}(\x,\y)\overline {\psi_{\r,\e,}(\y)}d\y.
\ea
Multiplying  (\ref{2-particle approx}) by $\overline \psi_{\r,\e}(\y)$, integrating over $B_L$ in $\y$, and using integration by parts yields} 
\beq\label{pre 1-equation B}
& &(-\nabla_x\cdotp \nabla_x+V_{\r,\e}(\x)+Q_\r(\x)-E)\,\Phi_L^{(1)}(\x)\\ \nonumber 
&=&2E_1\psi_{\r,\e}(\x)-
V_{eff}(\x)\psi_{\r,\e}(\x),
\eeq
with   the effective potential given by
\beq\label{effective potential}
V_{eff}(\x)=\int_{B_L} \frac 1{|\x-\y|}|\psi_{\r,\e}(\y)|^2 d\y.
\eeq
Next, we compare the above results with 
the case when a single particle scatters 
from the potential which is the sum of to the SH potential $V_{\r,\e}(\x)+Q_\r(\x)$
and the potential $V_{eff}(\x)$ multiplied by parameter $a$, that is the equation
\beq\label{Schr scattering Coulomb}
(-\nabla\cdotp \nabla+V_{\r,\e}(\x)+Q_\r(\x)+aV_{eff}(\x)-E^{eff})\,\phi^{eff}(\x)=0
\,\,\hbox{in }B_L,
\quad \phi^{eff}|_{\p B_L}=0.
\eeq
First-order perturbation theory then implies that  
\ba
& &E^{eff}=E+2aE_1+O(a^2)=E'+O(a^2),\\
& &\phi^{eff}(\x)=\psi_{\r,\e}(\x)+a\Phi_L^{(1)}(\x)+O(a^2)=\Phi_L(\x)+O(a^2).
 \ea 
 {\mmmmtext Thus, when  two particles are in a ball containing a SH potential,
 each particle behaves, up to error $O(a^2)$, as if the other particle
 and the SH potential were  replaced by the potential $V_{\r,\e}(\x)+Q_\r(\x)+aV_{eff}$.}
 }

{\mmmtext To analyze $V_{eff}$, let $\frak S$ be the strength of the SH potential, as defined in (\ref{11.06.11}).
Note that by choosing  $\tau$ appropriately,
we can make ${\frak S}$ arbitrarily large and, for  such  ${\frak S}$, the $L^2$ normalized
waves are strongly concentrated in  $B_{R}$. 
For  large ${\frak S}$ and every $\x \in B_L$,}
\beq\label{effective potential estimate}
\int_{B_L\setminus B_{R}} \frac 1{|\x-\y|}|\psi_{\r,\e}(\y)|^2 d\y<<
\int_{B_{R}} \frac 1{|\x-\y|}|\psi_{\r,\e}(\y)|^2 d\y.
\eeq

Recall that the potential $Q_\rho$ is constructed so that the solution
inside the cloak is concentrated in a ball $B_{R_0}$, and assume next that $R_0>0$ is small.
To emphasize this, we denote $R_0=\delta<<1$.
One then obtains
\beq\label{Q' formula}
V_{eff}(\x)\approx 
\int_{B_1} \frac 1{|\x-\y'|}|\psi_{\r,\e}(\y')|^2dy'\approx
 \frac {Q'}{|\x|},\quad Q'=\int_{B_1} |\psi_{\r,\e}(\y')|^2dy'
\eeq
{\ttext Thus,
$\Phi_L(\x)$ satisfies a one-particle Schr\"odinger equation
where the potential is a SH potential slightly modified  by a Coulomb
one with charge $aQ'$ at the origin.}

{\mmmtext Compare this with the case when we have no SH potential but
add the Coulomb interaction. Analyze this using   first-order perturbation theory,
 writing the wave function and the energy as 
 \ba
 & &\Phi^{Cou}(\x)=\phi^{empty}(\x)+a\phi_1^{Cou}(\x)+O(a^2),\\
  & &E^{Cou}= E+aE^{Cou}_1 +O(a^2),
 \ea
 where $ \phi^{empty}(\x)= c_L j_0(\omega_0|x|)$ and
  $c_L>0$ is such that $\|\phi^{empty}\|_{L^2(B_L)}=1$.
 Analogously to  (\ref{pre 1-equation B}), one obtains 
 \beq\label{Hemholtz scattering Coulomb simple no SH}
& &-\nabla\,\cdotp \nabla \phi_1^{Cou}(\x) -E^{Cou}\phi_1^{Cou}(\x)+a\V(\x) \phi_1^{Cou}(\x)=O(a^2)
,\quad\hbox{in }B_L\setminus 0,\\
& &\nonumber \phi_1^{Cou}|_{\p B_L}=0,
\eeq
where 
\ba
\V(\x)=\int_{B_L} \frac 1{|\x-\y'|}|\phi^{empty}(\y')|^2d÷y'
\ea
and
 \beq\label{derivative of energy no SH}
& & E^{Cou}_1=\frac 12 \int_{B_L}  
\int_{B_L} \frac 1{|\x-\y|}|\phi^{empty}(\x)|^2|\phi^{empty}(\y)|^2\,d\x d\y.
\eeq

Now compare $E_1$ defined for the Coulomb and SH potentials with the
analogous quantity
$ E^{Cou}_1$ defined for  the Coulomb potential (with no SH potential).
} Consider the case when ${\frak S}$ and $c_L$ are of the same size,
and when the potential $Q_\rho$ is constructed so that the solution
inside the cloak is concentrated in a ball $B_\delta$, where $\delta>0$
is small. Then formula (\ref{derivative of energy}) implies that we have  $E_1=O(\delta^{-1})$.
When $\delta$ is very small, the value $aE_1$, i.e., the change in
the energy level, is much larger than in the Coulomb case without SH potential.
Thus, by engineering the potential $Q_\rho$ appropriately, one
can make the SH cloak increase the interaction caused by the Coulomb
potential: When two particles are put
in a ball with a SH potential, the energy level of the particles is changed
as if the charges of the particles were multiplied by a factor
 of $O(\delta^{-1/2})$. The energy level coefficient $E_1$ then approaches  infinity as $\delta\to 0$,
{\mmmtext and thus for small $\delta$ one has that $E_1>>E^{Cou}_1$.}  
{This means that the presence of the SH potential has strengthened the Coulomb interaction
of the particles, effectively  increasing
repulsion between the particles. }

\section{Numerical simulations}
{\mmmtext  
{\bf Simulation of eigenfunctions and comparison with  free space.}
The results above  may be illustrated using numerical
 simulations.  For Fig. 3 we compute the effective field $\psi^{eff}$
 for the Dirichlet eigenfunctions 
in ball $B_L$ where in the ball have in $B_2$ the SH potential we use
$\rho=0.01,$ $L=2\pi$, $E=4$
and inside the cloak we have a potential $Q_0$ represented
in the form
\beq\label{Q_0 equation}
Q_0(r)=\tau_1\chi_{r<s_1}+\tau_2\chi_{s_1<r<s_2}.
\eeq
 using parameters
$s_1=0.6,$ $s_2=0.8$,
$\tau_2=-50$. Using Matlab, we found the value $\tau_1=12.9016$
corresponding to the SH potential. {\mmmmtext {\ttext  In the figures we will visualize the
effective field
 \beq \label{40b}
{\psi}^{eff}(\x)=\left\{\begin{array}{cl} \tilde\theta(\x)^\frac12 u(F^{-1} (\x)),&\hbox{for } 
\x\in B_L\setminus \overline B_1,\\
\beta u(0)\Phi(\x),&\hbox{for } \x\in \overline B_1, 
\end{array}\right.
\eeq
defined in formulas
 (\ref{33a}) and   (\ref{40a}), see also the formula (\ref{final convergence3}).}}

 Let  $R_a=3$ and $A_{empty}$ be the event
 ``the particle in the empty ball is in the layer $\{R_a<|x|<L\}$"
 and $A_{SH}$ be the event
 ``the particle in the  ball with the SH potential is in the layer $\{R_a<|x|<L\}$".
 Then 
 \ba
\prob(A_{empty})= 0.5021,\quad \prob(A_{SH})=    0.1355.
 \ea
 Additionally, let $B_{empty}$ be the event
 ``the particle in the empty ball is in the layer $\{2<|x|<L\}$"
 and $B_{SH}$  the event
 ``the particle in the  ball with the SH potential is in the layer $\{2<|x|<L\}$",
 i.e., the particle is outside the cloaking device. We obtain
  \ba
 \prob(B_{empty})= 0.7196,\quad \prob(B_{SH})=    0.1941.
 \ea
 The conditional probabilities that the particle is in $\{R_a<|x|<L\}$,
 conditioned on it being in the layer $\{2<|x|<R_L\}$, are given by
 \ba
 & & \prob(A_{empty}|B_{empty})= 
 \prob(A_{SH}|B_{SH})=   0.6977.
 \ea
{\mmmmtext This shows that when  the SH potential  is inserted
in the ball $B_L$, the particle is observed in the region $B_2-B_1$ with
a lower probability but, when it is observed, the observations from it are similar to
the observations one would have if the ball were empty.}

}

{\bf Numerical simulation of the scattered wave.}
{\mmmtext
We compute scattering solutions for the SH potential using spherical
harmonics $n\leq N=70$ and $|m|\leq n$. 
{\mmmmtext  In  Fig. 1 we plot the total
field corresponding to the plane wave with energy $E=256$
which scatters from the SH potential  supported in the ball $B_2$.
The figure shows the real part of the effective field  in
the ball $B_L$, $L=3$ in the plane $z=0$}. 
The SH potential corresponds to the parameter $\rho=0.01$ and} 
the potential $Q_0$
is represented in the form (\ref{Q_0 equation}) with 
$s_1=0.25$, $s_2=0.5,$ $\tau_2=-10$. Using Matlab, we found the
 value $\tau_1=-169.49$ corresponding to the SH potential.


%
%
}

{\mmmtext 
{\bf Simulation of different modes of the cloak.} We next  visualize the above choice of parameters using numerical
 simulations of the scattering problem. We compute
  the real part of effective field corresponding 
 to the total field when we have an incoming plane wave,
 {where in the ball  $B_2$ there is a potential consisting of a cloaking potential $V_{\r,\e}$
plus a  potential $Q_0$ supported in the cloaked region.}
We use   energy $E=4$, and  for the cloaking potential we use the parameter
 $\rho=0.01$  and the potential $Q_0$
 is given in the form  (\ref{Q_0 equation})
with paremeters
$s_1=0.6,$ $s_2=0.8$,
$\tau_2=-25$. Using Matlab, we found the value $\tau_1=7.0675$
corresponding to SH potential. We computed the corresponding
effective field, $\psi^{eff}_{SH}$, {\mmmmtext defined in  formula
(\ref{40b}).}
To show different modes of the cloak,
we perturbed the parameter $\tau_1$ to the value
$\tau_1^{cloak}=8.2531$
corresponding to the cloak mode and computed the corresponding
effective field  $\psi^{eff}_{cloak}$.
In addition, we perturbed the parameter $\tau_1$ to the value
value $\tau_1^{res}= 7.120$
corresponding to the resonance mode and computed the corresponding
effective field  $\psi^{eff}_{res}$. 
All fields are computed 
 using spherical
harmonics $n\leq N=30$ and $|m|\leq n$, 
and the real parts of the fields are plotted in Fig. 2
on the positive $x$-axis  $\{(x,0,0): \, x\in [0,3]\}$.
 }

\section{Implementation of the SH potentials}\label{sec: Implementation}
{\mmmtext 

{\mmmmtext Equation (\ref{equat 1C})
 with  isotropic mass $m_{\r,\e}$ and  bulk modulus $\kappa_R$
 describes  an approximate acoustic cloak. The negative values of $\tau_2$ 
 required  in our construction of potential $Q_\r$
corresponds then to a material with  a negative bulk modulus;
such materials have already been proposed \cite{Wang}.
There  are many proposals for acoustic cloaks
\cite{CummerSchurig,ChenChan,CummerEtAl,GKLU1};
implementing an acoustic cloak and placing negative bulk modulus
material inside the cloak, one could test the concept of the SH potential in the acoustic setting.

For electromagnetic waves, one can consider a cylindrical  electromagnetic cloak 
\cite{SchurigEtAl} and insert  in it material with negative permittivity
to implement a structure similar to the SH potential.
The analysis related to such cylindrical cloaks with suitable chosen
parameters to create a SH potential for incident
TM-polarized waves is similar to the analysis
for the 3D cloak considered in this paper {\ttext (with different asymptotics of coefficients (\ref{8.5.2})).}

Next we consider approximate quantum cloaks and Schr\"odinger hat potentials.}

To implement an anisotropic Schr\"odinger hat, one could consider a quantum 
cloak with an anisotropic effective mass and include the potential $Q_\rho$ in the 
cloaked region. Possible realizations of quantum cloaks have been proposed by 
{Zhang} et al. \cite{Zhang}, using  crystal structures in an optical lattice  at ultra-low temperatures,
which make possible a large variation of the effective mass. Below, using 
the Liouville gauge transformation (\ref{basic gauge}), we 
propose a realization of the SH potential using 
solid state models which do not require large variation of the effective mass. 

The function $V_{\r,\e}$ is a rapidly oscillating radially symmetric
potential, and  {\mmmtext we can use the homogenization theory to approximate the SH potential $V_{\r,\e}+Q_\r$ 
by a piecewise} constant radially symmetric function,
$$
V_{SH}({\bf r})=\sum_{j=1}^{N} V_j\chi^{(1)}_j(|{\bf r}|),\quad {\bf r}\in \R^3,
$$
where $\chi_j^{(1)}$ are indicator functions of  suitably chosen 
intervals $I_j^{(1)}=[a(j),b(j)]$ {\ttext and $V_j$ are constants}. 
Applying  homogenization theory  {\mmmmtext one can see that the solutions
corresponding to $V_{SH}$ with  sufficiently large $N$ approximate the solutions corresponding to $V_{\r,\e}+Q_\r$
when $V_j$ are chosen appropriately. In fact, it is enough to use 
potentials for which} the  constants $V_j$ have only
two different values, one very negative and one very positive. Thus, we can  form an approximate SH potential
using layers of
spherical potential wells of  depth $-V_-$ and barrier walls of height $V_+$, i.e.,
$$
V_{SH}({\bf r})=\sum_{j=1}^{N_1} V_+\chi_j^+(|{\bf r}|)-\sum_{j=1}^{N_2} V_-\chi_j^-(|{\bf r}|)
$$
where $\chi_j^\pm$ are indicator functions of the suitably chosen 
intervals $I_j^\pm=[a^\pm(j),b^\pm(j)]$.
Note that the two-scale homogenization theory used for the above approximation
does not specify how large $N$ one needs to use, but just states that the convergence
to the correct limit happens as $N$ grows.

Let us choose the physical units so that  $m_0=1$ is the effective mass of the particle which we consider and
that $\hbar=2^{1/2}$.
If $\psi$ satisfies
$$
(-\frac{\hbar^2}{2m_0}\nabla\,\cdotp \nabla+V_{SH}({\bf r})- E)\psi({\bf r})=0,
$$
then by scaling the length variable by $\ell$, we see that $\Psi({\bf r})=\psi({\bf r}/\ell)$ satisfies
\beq\label{small amplitude cloaking potential}
(-\frac{\hbar^2}{2m_0}\nabla\,\cdotp \nabla+\ell^{-2}  V_{SH}({\bf r}/\ell)-\ell^{-2}  E)\Psi({\bf r})=0.
\eeq
When $\ell$ is very large,
this means that the energy level $ E$ is replaced by a new,  much smaller energy level 
$\ell^{-2}  E$ and the depth of the potential wells are replaced by much  shallower 
 wells, and the height of the barrier walls are replaced by much  lower
 walls.
We denote
$
V_{SH,\ell}({\bf r})=\ell^{-2}  V_{SH}({\bf r}/\ell)$ and $E_\ell=\ell^{-2}E$.
When the above potential $V_{\r,\e}+Q_\r$ is defined using energy $E$ in formulas (\ref{Vre}) and (\ref{Qrho}),
we say that $V_{SH,\ell}$ is designed to operate at the  energy level $E_\ell$.

{To consider models appearing in solid state physics, we have to consider an effective mass
depending on the $x$ variable.}
So,  now consider four materials with isotropic (i.e., spherical) effective masses $m_1,m_2,m_3,m_4$,
 such that $m_1\leq m_2\leq m_3\leq m_4$ and  $m_1\leq m_0\leq m_4$ 
and potentials $V_1,V_2,V_3,V_4$ corresponding to the conduction band edge energies.
 Assume that  the maximum of $V_j$ is larger than the maximum $V_{SH,\ell}$ and
the minimum of $V_j$ is smaller than the minimum $V_{SH,\ell}$.
We next study  spherical layers of these materials,
with abrupt interfaces between materials.

Let us consider a structure consisting of   many thin spherical layers of these four materials which
 near $|{\bf r}|=r$ are  
mixed according to ratios $\ell_1(r),\ell_2(r),\ell_3(r),\ell_4(r)\in [0,1]$, correspondingly. We need 
these ratios to satisfy the system
\beq\label{eq: Homogenized}
& &\ell_1(r)+\ell_2(r)+\ell_3(r)+\ell_4(r)=1,\\ \nonumber
& &m_1\ell_1(r)+m_2\ell_2(r)+m_3\ell_3(r)+m_4\ell_4(r)=m_0,\\ \nonumber
& &\frac {\ell_1(r)}{m_1}+\frac {\ell_2(r)}{m_2}+\frac {\ell_3(r)}{m_3}+\frac {\ell_4(r)}{m_4}=\frac {1}{m_0},\\ \nonumber
& &V_1\ell_1(r)+V_2\ell_2(r)+V_3\ell_3(r)+V_4\ell_4(r)=V_{SH,\ell}({\bf r}).
\eeq
Below, we assume  that $m_j$, $V_j$, $E_\ell$, and $\ell^{-2}V_\pm$ are such that the solution
of the system (\ref{eq: Homogenized}) satisfies  $\ell_1(r),\ell_2(r),\ell_3(r),\ell_4(r)\in [0,1]$.
{By means of  homogenization theory, one can see that when}
the equations
(\ref{eq: Homogenized}) hold,
we can approximate the SH potential 
using a configuration where spherical layers of materials $(m_j,V_j)$ 
are combined at the energy level $E_\ell$, as is shown below. Note that if the masses $m_j$ are equal to $m_0$ we need
only two materials and approximate the SH potential with spherical quantum wells of given depth $-\ell^{-2}V_-$ and walls of height $\ell^{-2}V_+$.
In the general case when the masses are different, we need four different materials to solve equations
(\ref{eq: Homogenized}). 

Now consider the situation where 
we have spherical layers
\ba
L_j=B_{R(j)}-B_{R(j-1)}\subset \R^3,\quad j=1,2,\dots,4J
\ea
with $0\leq R(j)<R(j+1)\leq 2$.
 Let $\psi$ be the wave function corresponding to the particle in this layered
structure. Then in the each layer we have Schr\"odineger
equation
\beq\label{model 1A}
(-\frac {\hbar^2}{2}\nabla \cdotp\frac 1{m_{i}} \nabla +V_i-E_\ell)\psi({\bf r})=0,\quad x\in L_j,\quad i=i(j)
\eeq
and $i(j)\in \{1,2,3,4\}$ indicates which material is present in the layer $L_j$.
We may choose $i(j)\equiv j$ mod $4$ and 
\ba
R(j+1)-R(j)=\frac 1{2J}\ell_{i(j)}(R(j)).
\ea
On the interfaces of the layers, that is, at $r=R(j)$, we impose the BenDaniel-Duke boundary
conditions \cite{BD-D} 
\beq\label{model 1B}
\psi|_{r=R(j)+}=\psi|_{r=R(j)-},\quad  \frac 1{m_{i(j+1)}}\p_r \psi|_{r=R(j)+}=  \frac  1{m_{i(j)}}\p_r \psi|_{r=R(j)-}.
\eeq
Similar Hamiltonians and the appropriate boundary conditions in various
heterostructures are discussed extensively {\ttext in the reference [\onlinecite{M}].}
  When number of layers $4J>>N$, the solutions of the obtained equations approximate
the solutions of the equation (\ref{small amplitude cloaking potential}),
on related mathematical theory, cf.  [\onlinecite{A-P,GKLU_JST}].

We now propose how the above model could be physically
implemented using  effective-mass theory to approximate
electrons in semiconductors. Consider a semiconductor heterostructure build up using 
four semiconductor materials
having the same lattice structures, such as $\hbox{Al}_x\hbox{Ga}_{1-x}\hbox{As}$ with  mixing
parameter $x$ having the values $x_1$, $x_2$, $x_3$, and $x_4$ where $0\leq x_j\leq 1$,
and assume that the lowest energies in each conduction band (the conduction band edge)
in these materials corresponds to the wave vector $\k_0=0$.
Using Bastard's
envelope function approximation \cite{B,Bastard}, 
we consider in the heterostructure the wave function $\psi({\bf r})$,  corresponding to
the energy $\E$ being close to a  conduction band edge,  which  in each material can be expanded as 
\beq\label{envelope functions}
\psi({\bf r})=\sum_n f_n({\bf r}) u_n({\bf r},\k_0)
\eeq
where the sum in $n$ is taken over the finite number of energy bands. Here, $f_n({\bf r})$
are the slowly varying envelope functions and $u_n({\bf r},\k_0)$ are the periodic Bloch functions
in the material corresponding to wave vector $\k_0$. We also assume that the Bloch functions 
are the same in all four materials.
Next we consider  the single band analysis of electrons near the lowest
conduction band energy
 and study
the wave function $\psi({\bf r})$ 
omitting  in the sum (\ref{envelope functions}) all other values of $n$ except the value $n_0$ corresponding to the lowest energy in the conduction band.
{\ttext  This means that we use the approximation
\beq\label{envelope functions single band}
\psi({\bf r})= f_{n_0}({\bf r}) u_{n_0}({\bf r},\k_0).
\eeq

For the $\hbox{Ga}\hbox{As}$-$\hbox{Ga}(\hbox{Al})\hbox{As}$ heterostructures, with 
sufficiently thick $\hbox{Ga}\hbox{As}$ layers,  the above approximation (\ref{envelope functions single band}),
with the Schr\"odinger equations
\beq\label{model 1A for f}
(-\frac {\hbar^2}{2}\nabla \cdotp\frac 1{m_{i}} \nabla +V_i-\E) f_{n_0}({\bf r})=0,\quad {\bf r}\in L_j,\quad i=i(j),
\eeq
and
the BenDaniel-Duke interface conditions (\ref{model 1B}) for the single band envelope function $f_{n_0}({\bf r})$,
 have been proposed in  [\onlinecite{Bastard},\ Sec. 3.II.2.3] and [\onlinecite{BB}]. Similar models
 have also been proposed for other 
3D heterostructures in \cite{BD-D,M}}.  


Now suppose that the  spherically layered
semiconductor structure described above is  located
in the ball $B_2$;  surround it with a similar
heterostructure having 
effective mass $m_0$ and the conduction band edge energy $V_0$, being normalized to have the value $V_0=0$.
We note that above we could have assumed that, e.g., \ $m_2=m_0$ and $V_2=0$ in which case the surrounding material
may be homogeneous semiconductor material. 
To consider the scattering of electrons traveling in the medium having the energy $\E$,
one can study the Dirichlet-to-Neumann operator $\Lambda:h\mapsto \p_\nu f_{n_0}|_{\p B_2}$ for the boundary value problem,
%
%
\beq\label{DN PROBL}
& &(-\frac {\hbar^2}2\nabla\,\cdotp \frac 1{m({\bf r})}\nabla+V({\bf r})
)f_{n_0}({\bf r})=\E f_{n_0}({\bf r}),\quad \hbox{for }{\bf r} \in B_2,\\
& &f_{n_0}({\bf r})|_{\p B_2}=h, \nonumber
\eeq
see (\ref{Omega SH})-(\ref{Omega SH, hole}).
Here, $m({\bf r})=m_{i(j)}$ and $V({\bf r})=V_{i(j)}$ in  spherical layers $L_j\subset B_2$. 

As the number of layers, $J$, grows, the envelope function
 $f_{n_0}({\bf r})$ approaches \cite{A-P,GKLU_JST} the solution of the Schr\"odinger equation 
 $
(-\frac{\hbar^2}{2m_0}\nabla\,\cdotp \nabla+ V_{SH,\ell}-\E)f=0
$
 and, moreover, the Dirichlet-to-Neumann map $\Lambda$ of the equation (\ref{DN PROBL}) approaches the Dirichlet-to-Neumann operator corresponding to the equation
\ba
& &-\frac {\hbar^2}2 \nabla\,\cdotp \frac {1}{m_0}\nabla f^{hom}({\bf r})=\E f^{hom}({\bf r}),\quad \hbox{for }{\bf r} \in B_2
\ea
corresponding to the homogeneous background.
 Thus, the scattering of electrons caused by the heterostructure  
in $B_2$ is very small but the wave function may be very large inside the ball $B_1$.
We emphasize that $V_{SH}$ depends on the energy level $E_\ell$, see (\ref{Vre}) and (\ref{Qrho}), and thus
the above analysis applies only for electrons whose energy $\E$ is close to $E_\ell$.} 
%

We note that the above theoretical model can be considered
as a (much) more complicated structure than the 
spherical semiconductor 
layer construction previously  used to implement  quantum dots \cite{SMEW}
and a related construction of cylindrical semiconductor
layers \cite{TPM}.

{




Let us also discuss the distribution of the energies of electrons in
the heterostructure with conduction band edge energy $\E_c$. The  density function of the energies of the electrons in the  conduction band
is the product
$$
n(\E)=D(\E)f(\E)
$$
of the density of states $D(\E)$ and the Fermi-Dirac distribution 
$f(\E)=C(e^{(\E-\mu)/(k_BT)}+1)^{-1}$,
where $T$ is the temperature, $\mu<\E_c$ is the Fermi energy,  $k_{B}$ is Boltzmann's
constant, cf. [\onlinecite{Sze}, Sec.\ 1.4] and [\onlinecite{Kittel}, Sec.\ 8], and $C$ is  the normalization constant. The density of  states $D(\E)$ is usually
approximated by $C(\E-\E_c)^{1/2}$, $\E>\E_c$ near $\E_c$. When
the Fermi-Dirac distribution is approximated by Maxwell-Boltzmann
distribution  $f_{mb}(\E)=Ce^{-(\E-\mu)/(k_BT)}$,
 one sees that the density
of the energy $\E-\E_c$ is approximately distributed according
to the Gamma distribution with the shape parameter $3/2$ and
the scale parameter  $k_BT$. Then the energies
have the expectation
$$
\E_{av}=\E_c+\frac 32 k_BT
$$
and the variance  $\frac 32  (k_B T)^2$. Thus,
the energies near the conduction band edge $\E_c$ are crucial
in the modeling of semiconductors, and at low temperatures $T$ the
density of electrons, $n(\E)$, is concentrated near the average $\E_{av}$.


By the above, when the temperature $T$ is low enough, then  
the energies of the electrons in the conduction band have a distribution
concentrated near the energy level $\E_{av}$. Thus by perturbing a
homogeneous semiconductor material by including in it one
or several SH potentials $V_{SH,\ell}$, designed to operate at the  energy level  $E_\ell=\E_{av}$, 
one can create a device where most of the electrons would behave
in the above analysis: The non-normalized wave functions of electrons  would not
be perturbed outside the supports of the SH potentials but the amplitude of
the wave functions are strongly
amplified inside the support of the SH potentials. 
}

In summary: 
Assume that we have layers of
semiconducting materials where the  electrons {\ttext with energy near the edge of the conduction
band can be modeled by
a Schr\"odinger equation (\ref{DN PROBL})}  and
that in these semiconducting materials the effective masses 
and the potentials have slightly different values.
Then, appropriate choices of  layer thickness  and a sufficiently large heterostructure  lead to} 
the envelope functions of the wave functions satisfing  the Schr\"odinger equation for a SH potential. 
}

{\bf Acknowledgments:} AG is supported by  US NSF; YK by  UK EPSRC; ML by Academy of Finland; and GU by  US NSF, a Walker Family Endowed Professorship at UW, a Chancellor Professorship at UC, Berkeley, and a Clay Senior Award.

\end{document}